\documentclass[]{aastex631}

\usepackage{amsmath}
\usepackage{tabularx}
\usepackage{array}

\newcommand{\degree}{\ensuremath{^\circ}}

\newcommand{\p}[1]{{\color{magenta}{#1}}}

\newcommand{\bs}[1]{{\color{orange}{#1}}}

\shorttitle{Impact of solar activity on ICMEs}
\shortauthors{Perri et al.}

\graphicspath{{./}{figures/}}

\begin{document}

\title{Impact of the solar activity on the propagation of ICMEs: \\ Simulations of hydro, magnetic and median ICMEs at minimum and maximum of activity} 

\correspondingauthor{Barbara Perri}
\email{barbara.perri@universite-paris-saclay.fr}

\author[0000-0002-2137-2896]{Barbara Perri}
\affil{Université Paris-Saclay, Université Paris Cité, CEA, CNRS, AIM, 91191, Gif-sur-Yvette, France}
\affil{Centre for mathematical Plasma Astrophysics, KU Leuven, 3001 Leuven, Belgium}

\author[0000-0003-3364-9183]{Brigitte Schmieder}
\affil{Centre for mathematical Plasma Astrophysics, KU Leuven, 3001 Leuven, Belgium}
\affil{LESIA, Observatoire de Paris, France}

\author[0000-0001-8215-6532]{Pascal Démoulin}
\affil{LESIA, Observatoire de Paris, France}
\affil{Laboratoire Cogitamus, rue Descartes, 75005 Paris, France}

\author[0000-0002-1743-0651]{Stefaan Poedts}
\affiliation{Centre for mathematical Plasma Astrophysics, KU Leuven, 3001 Leuven, Belgium}
\affiliation{Institute of Physics, University of Maria Curie-Sk{\l}odowska, Pl.\ M.\ Curie-Sk{\l}odowskiej 5, 20-031 Lublin, Poland}

\author[0000-0002-4017-8415]{Florian Regnault}
\affil{Space Science Center, University of New Hampshire, Durham, New Hampshire, United States}

\begin{abstract}

The propagation of Interplanetary Coronal Mass Ejections (ICMEs) in the heliosphere is influenced by many physical
phenomena, related to the internal structure of the ICME and its interaction with the ambient solar wind and magnetic field. As the solar magnetic field is modulated by the 11-year dynamo cycle, our goal is to perform a theoretical exploratory study to assess the difference of propagation of an ICME in typical minimum and maximum activity
backgrounds.
We define a median representative CME at 0.1~au, using both observations and numerical simulations, and describe it using a spheromak model. 
We use the heliospheric propagator European Heliospheric FORecasting Information Asset (EUHFORIA) to inject the same ICME in two different background wind environments. 
We then study how the environment and the internal CME structure impact the propagation of the ICME towards Earth, by comparison with an unmagnetized CME. 
At minimum of activity, the structure of the heliosphere around the ecliptic causes the ICME to slow down, creating a delay with the polar parts of the ejecta. This delay is more important if the ICME is faster. At maximum of activity, a southern coronal hole causes a northward deflection. For these cases, we always find that the ICME at maximum of activity arrives first, while the ICME at minimum of activity is actually more geo-effective. The helicity sign of the ICME is also a crucial parameter but at minimum of activity only, since it affects the magnetic profile and the arrival time of up to 8 hours.

\end{abstract}


\section{Introduction} 
\label{sec:intro}
Space weather is the ability to anticipate sudden events from the Sun and their impact on our planet \citep{Schrijver2015_cospar}. Coronal Mass Ejections (or CMEs) are considered one of the main drivers of strong space weather events \citep{Gosling1993}: they consist of large-scale ejections of plasma and magnetic field by the Sun, which then travels through the inner heliosphere until they eventually reach our planet \citep{Webb2012}. CMEs that have an impact on Earth are called geo-effective \citep{Koskinen2006}. In particular, when they interact with the Earth magnetosphere, they can generate magnetic storms that have consequences for our atmosphere (polar lights) and ground (induced currents) \citep{Pulkkinen2007}. CMEs are an important concern for space weather due to their frequency (more than 10 times a day launched in all directions during intense phases of solar activity, see \cite{Robbrecht2009}), and their impact on our technology (severe electrical damages due to induced currents, see \cite{Pirjola2005}). The most powerful recorded event was the Carrington event on 1st of September 1859, where a CME traveled at 2300 km/s and reached Earth within 17 hours \citep{Tsurutani2003}; the cost of the impact of such an event happening nowadays has been evaluated at trillions of dollars \citep{Schrijver2015}.

CMEs originate mostly from active regions, which are regions where the magnetic field is particularly intense, and stored in sheared and twisted structures (usually flux-ropes, see \cite{Demoulin2008,Schmieder2015}). Eventually, these structures become unstable, and the plasma trapped inside is released into the heliosphere as magnetic ejectas \citep[MEs,][]{Winslow2015}.  When observed close to the Sun, most CMEs are characterized with a bright frontal loop, a dark cavity, and an embedded bright core \citep{Illing1985} possibly corresponding to the erupting filament \citep{House1981}. Images are available using white-light images via Thomson scattering to follow their initial propagation \citep{Davies2013}. The in situ counterpart of CMEs have been historically called Interplanetary Coronal Mass Ejections (ICMEs). Their most important components are both hydrodynamic and magnetic \citep{Dumbovic2015}. The speed and density of the ICME, contribute to the dynamic pressure and also contain signatures of the propagation of the interplanetary shock at the front of the ICME. Interplanetary shocks alone can drive geomagnetic activity \citep{Oliveira2018}, which is why the most simple models do not include an internal magnetic field structure for the ICME. However, it has been shown that it is its magnetic field amplitude and orientation which are driving the strongest geomagnetic storms \citep[especially its $B_z$ component which favors dayside reconnection with the magnetopause,][]{Lugaz2016}. The interplanetary shock is followed by a compressed and heated region called the sheath \citep{Kilpua2017}. This region is caused by the accumulation of solar wind at the front of the ICME, as well as the expansion of the following ME \citep{Kaymaz2006}. An ME is characterized by a strong and smooth magnetic field, as well as low temperature and low plasma beta \citep[ratio of the thermal pressure over the magnetic pressure,][]{Wang2005}. Other diagnostics can also be used when available \citep{Zurbuchen2006}. When the magnetic ejecta shows rotation of its magnetic field, and has a proton temperature lower by a factor two than the typical solar wind with the same speed, it is categorized as a magnetic cloud (MC), as it is associated with the existence of a flux-rope \citep{Burlaga1981, Burlaga1995}. Such configuration only happen for about one third of ICMEs at 1~au \citep{Wu2011}. However, this may be due to the limitation of having only one measurement point \citep{Jian2006, Kilpua2011}. The recent era allows for multi-spacecraft coordination thought the heliosphere to try to better quantify the 3D geometry of ICMEs \citep{Mostl2022}.

As ICMEs propagate through the heliosphere, they interact with the various structures they encounter, and as a result evolve. For example, ICMEs will naturally expand as they travel, and usually it can be approximated by a self-similar expansion \bs {\citep{Demoulin2008expansion,Gulisano2010,Gulisano2012,Chane2021,Verbeke2022}}. Concerning their trajectory, CMEs can suffer deflections both in latitude and longitude, because of their interaction with the magnetic field of coronal holes, helmet streamers and the heliospheric current sheet (HCS) \citep{Gopalswamy2014}. This tends to focus CMEs towards the Earth latitudes (especially at minimum of activity, see \cite{Zuccarello2012}).
Recent studies have shown that the strength and sign of the ambient magnetic field can influence their drift \citep{Asvestari2022} due to tilting instabilities \citep{Bellan2000}. It is also linked to the sign of helicity of the source region \citep{Green2007}. Concerning their speed, ICMEs can be accelerated or decelerated through their interaction with the ambient solar wind \citep{Gopalswamy2000} which causes drag-like effects \citep{Cargill1995}. Concerning their magnetic flux, it can be reduced due to magnetic reconnection, which will lead to magnetic erosion at the front of the ICME \citep{Dasso2007, Ruffenach2012}. As they propagate, ICMEs are more likely to become more and more complex as a result of their interaction with solar wind structures 
\citep{Winslow2015,Winslow2022,Scolini2022,Scolini2023}, and to display aging processes that will contribute to their deformation \citep{Demoulin2020}. They will also change as a result of interactions with specific structures, such as high-speed streams (HSSs) \citep{Fenrich1998, Heinemann2019, Scolini2021_hss}, stream-interaction regions (SIRs) or other ICMEs \citep{Lugaz2005, Scolini2020}. For more details, see reviews by \cite{Lavraud2014} and \cite{Shen2022}, and references within.

The medium in which the ICMEs propagate is far from simple to describe, as it shows great complexity and variability. The interplanetary medium is influenced both by the magnetic field close to the Sun, and then the solar wind further away from it. The solar magnetic field is generated inside the Sun by dynamo effect \citep{Moffatt1978, Parker1993, Brun2017}, and then bathes the entire heliosphere following the Parker spiral due to the rotation of the star \citep{Owens2013}. The solar wind is made of plasma particles continuously emitted by the Sun \citep{Parker1958}. It has two main components, one slow and one fast, whose source mechanisms differ (see \cite{Cranmer2007}, and \cite{Viall2020} and references within). The short-term variability of the interplanetary medium is caused by reconnection effects in the lower corona (causing the switchbacks observed by Parker Solar Probe, see \cite{Kasper2019}), and transient events (perturbations caused by SIRs, HSSs, previous ICMEs...). The long-term variability on the other hand is due to the solar activity cycle \citep{Hathaway2015}. Indeed, the solar dynamo is cyclic with a period of 22 years \citep{Weiss1990}, which generates periods of low magnetic activity, called minima, where the field is mostly dipolar, and periods of high magnetic activity, called maxima, where the field becomes more multipolar \citep{Hoeksema1984, Derosa2012}. This modulation has a direct effect on the structure of the corona and inner heliosphere. At minimum of activity, the corona is very structured with stable equatorial helmet streamers, thus confining the slow solar wind to the equator and the fast wind at the poles; at maximum of activity, the corona is more complex with pseudo-streamers and streamers emerging at various latitudes, and thus the solar wind has slow and fast streams at all latitudes \citep{McComas2003, McComas2008}.

Because of all these complex interactions, the impact of solar activity on ICME propagation is still unclear. We know that CMEs/ICMEs are more frequent \citep{Gopalswamy2003}, faster \citep{Hundhausen1990, Dasso2012} and more magnetized \citep{Wu2011} at maximum of activity, and yet the most powerful events recorded (like the Carrington event for example) did not necessarily happen during these periods (because they are probably due to ICME-ICME interactions) \citep{Chapman2023}. There are other interesting relationships between ICMEs and the solar cycle. The sign of the helicity can be estimated using hemispheric rules that depend on the solar cycle polarity \citep{Bothmer1998}. \cite{Gopalswamy2003} noted that high-latitude CMEs do not occur during polarity reversals phases of the solar cycle, due to the lack of closed field lines close to the poles. Studies have analyzed the dependence of specific properties with the phase of the solar cycle. \cite{Dasso2012} showed that the expansion of the ICME is not related to the solar cycle, but the amount of helicity of the magnetic cloud is. \cite{Jian2011} showed that during solar minima, shocks are more important due to a slower solar wind. Finally, \cite{Regnault2020} analyzed 20 years of ACE data and compared the internal properties of ICMEs based on the solar activity phase: they confirmed that ICMEs during maximum were faster, but otherwise parameters were similar; what changes is that at maximum, they observe a larger distribution of ICME parameters, just like in \cite{Wu2016} where they used Wind data.

Our goal with this paper is to explore more quantitatively the impact of solar activity on the propagation and geo-effectiveness of ICMEs.  In this first exploratory study, we will inject the same ICME in two different activity background using numerical simulations, and quantify the differences and their origins, as a first step towards understanding what effects to expect.

The article is organized as follows. In Section \ref{sec:euhforia}, we describe the numerical set-up behind the European Heliospheric FORecasting Information Asset (EUHFORIA) code, for both the coronal and heliospheric part, as well as the CME/ICME modeling. In Section \ref{sec:bc}, we explain our choice of boundary conditions, first for the solar wind, then for the CME insertions. In Section \ref{sec:limit_cases}, we present two limit cases: a first one with only solar wind to quantify the background for the next study of ICME propagation, and a second one where the injected CME is purely hydrodynamic, to analyze the effect of the solar wind speed. In Section \ref{sec:cme_12_july}, we present the results for a magnetized CME inspired by a real event. 
In Section \ref{sec:cme_median}, we present the results for a ICME representing a median case based on statistical analysis of ICME parameters. Finally, in Section \ref{sec:conclusion}, we present our discussion and conclusion.

\section{Description of EUHFORIA simulations} 
\label{sec:euhforia}

We use the numerical 3D MHD code EUHFORIA \citep{Pomoell2018}. It is divided into two numerical domains with different treatments: a semi-empirical model for the coronal part (from 1 $R_\odot$ to 21.5 $R_\odot$), and an MHD model for the heliospheric part (from 0.1 to 2~au). The limit between the two domains is set at 0.1~au because at this distance the solar wind is superfast, making the coupling one-way (the heliospheric part cannot back-react on the coronal part).

\subsection{Coronal part}
The coronal part uses synoptic magnetic maps as inputs, which are observations of the photospheric radial magnetic field used to drive the simulation. They are called synoptic because they cover the 360$\degree$ of the solar surface, but they are not necessarily synchronic (which means that the solar observations displayed on the map were not necessarily taken at the same date, for example the observations can last a full solar rotation and thus have a 27-day gap) \citep{Riley2014}. The default set-up of EUHFORIA can use two types of magnetic maps: the ones from the Global Oscillation Network Group (GONG) \citep{Harvey1996} and the ones from the GONG Air Force Data Assimilative Photospheric Flux Transport Model (GONG-ADAPT) \citep{Arge2010, Hickmann2015}. From this input, the magnetic field global configuration is derived using two models. First, we use a potential-field source surface (PFSS) model up to the source surface $R_{ss}$ of 2.6 $R_\odot$ \citep{Altschuler1969}. This allows us to compute a current-free configuration by assuming the magnetic field is potential below the source surface, and purely radial afterwards. Then it is coupled to a Schatten current sheet (SCS) model from 2.3 $R_\odot$ up to 21.5 $R_\odot$ \citep{Schatten1969}. The SCS model starts slightly below the source surface to reduce possible kinks due to incompatibility between the two models \citep{McGregor2008}. The SCS field is required to vanish at infinity, in order to extend the magnetic field radially while maintaining a thin structure for the heliospheric current sheet (HCS). This method provides results in better agreement with observations (with the Ulysses mission, for example, see \cite{Pinto2017}). 

Once the magnetic configuration is computed, a semi-empirical Wang-Sheeley-Arge method is used to compute the radial velocity \citep{Wang1990, Arge2003}, using the following formula (in km/s): 
\begin{equation}
    v_r(f,d) = 240 + \frac{675}{(1+f)^{0.222}}\left[1.0 - 0.8\,\rm{exp}\left(-\left(\frac{d}{0.02}\right)^{1.25}\right)\right]^3,
    \label{eq:wsa}
\end{equation}
where $d$ is the angular distance of the footpoint of the magnetic field line to the closest coronal hole boundary, and $f$ is the flux tube expansion factor between the photosphere and the source surface. This law is adapted from \cite{vanderHolst2010, McGregor2011}. Since the solar wind continues to accelerate beyond 0.1~au, we deduct a constant value of 50 km/s from equation (\ref{eq:wsa}) to prevent the wind speed from being systematically overestimated. Furthermore, we enforce the final speed to be in the range $v_r\in[275, 625]$ km/s by capping it \citep{McGregor2011}. Finally, we apply a rotation of 10$\degree$ to the obtained solar wind speed map at 0.1~au to take into account the approximated solar rotation that is not included in the magnetic field model.

Based on this computed velocity, the density $n$, temperature $T$ and radial magnetic field $B_r$ are computed at 0.1~au using the following relations \citep{Pomoell2018}:
\begin{equation}
    n = n_{fsw}\left(v_{fsw}/v_r\right)^2,
\end{equation}
\begin{equation}
    T = T_{fsw}\left(\rho_{fsw}/\rho\right),
\end{equation}
\begin{equation}
    B_r = \rm{sgn}\left(B_{corona}\right)B_{fsw}\left(v_{r}/v_{fsw}\right),
\end{equation}
with $v_{fsw}=675$ km/s, $n_{fsw}=300 \, \rm{cm}^{-3}$, $T_{fsw}=0.8$ MK, $\rho_{fsw}=0.5n_{fsw}m_p$ ($m_p$ being the proton mass), $B_{sfw} = 300$ nT and $\rm{sgn}\left(B_{corona}\right)$ being the sign of the magnetic field 
as given originally by the coronal model at 0.1~au.
All these values are consistent with a fast solar wind (hence the abbreviation $fsw$) \citep{Odstrcil1999}. The number density prescription ensures a constant kinetic energy density on the spherical surface at 0.1~au. The plasma thermal pressure is chosen to be constant at 0.1~au (equal to 3.3~nPa), which sets the fast wind temperature. The reconstruction of the magnetic field based on the radial speed instead of the PFSS+SCS avoids the open-flux problem denoted by \cite{Linker2017}. These conditions are similar to the ones described in \cite{Odstrcil1999}.

\subsection{Heliospheric part}
Using the boundary conditions provided by the coronal part, the heliospheric part of EUHFORIA then computes the solar wind all the way to 2~au by solving the 3D time-dependent ideal MHD equations augmented with gravity. The equations are solved in the heliocentric Earth equatorial (HEEQ) frame, which is described as the frame with its Z-axis aligned with the rotation axis of the Sun, and its X-axis defined by the intersection of the solar equatorial plane and the solar central meridian of date as seen from the Earth. Although the chosen frame is not inertial, we choose to omit the Coriolis and centrifugal terms which should be the result of the orbital motion of Earth, as their contribution is actually negligible. A value of 1.5 is selected for the polytropic index as in \cite{Odstrcil2004}. This value is close to the ones found for protons in all solar winds with Helios and Parker Solar Probe \citep{Dakeyo2022}. The fact that we use a reduced index closer to 1 than 5/3 is a simple way of modelling coronal heating and as a result acceleration of the solar wind
\citep{Pomoell2012}.

A relaxation phase is first performed (set to 14 days in our case to avoid any spurious effect caused by the initial transient at the outer boundary condition). Then the forecast phase begins at the date set by the magnetic map. The forecast phase last 7 days in our cases, which is more than enough for the ICME to cross the entire domain. 

The default set-up has a 2$\degree$ angular resolution. The computational domain extends from 0.1 to 2~au in the radial direction and spans 120$\degree$ in latitude and 360$\degree$ in longitude. This means that the solar poles are truncated by 30$\degree$ in each hemisphere. To solve the MHD equations, we use a finite volume method combined with a constrained transport approach. To obtain a scheme which is both robust and second-order accurate, we use an approximate Riemann solver with standard piece-wise linear reconstruction \citep{Kissmann2012, Pomoell2012}. At the outer radial boundary, we use open boundary conditions implemented via a simple extrapolation, whereas at the latitudinal boundaries we use symmetric reflection boundary conditions.

One last point of detail we would like to discuss here is the difference between EUHFORIA outputs and proton data. As explained in \cite{Scolini2021_radial}, EUHFORIA uses a single-fluid approach, and thus makes no difference between the different particle populations (such as protons, electrons, $\alpha$ particles, etc.). In order to compare EUHFORIA outputs to observational data, we must thus make some assumptions. We will consider the proton and electron populations as the two primary contributors to solar wind plasma, and we assume that the two species have the same temperature: $T = T_p = T_e$. To further ensure the
quasi-neutrality of the plasma at all locations and times in the heliosphere, we also assume the two species have the same number density: $n_p=n_e$. This means that the plasma density in EUHFORIA is $n=n_p+n_e=2n_p$. As a result, EUHFORIA thermal pressure and $\beta$ plasma parameter is also twice that of the proton population: $P_{th} = P_{p,th}+P_{e,th} = n_pk_BT_p + n_ek_BT_e = 2P_{p,th}$ and $\beta=P_{th}/P_{mag}=2P_{p,th}/P_{mag}=2\beta_p$. The plasma total pressure is then computed as: $P_{tot} = P_{mag}+P_{th} = P_{mag}+2P_{p,th}$. The plasma temperature is retrieved using: $T=T_p=P/nk_B=P_p/n_pk_B$. Since $m_p \gg m_e$, we can assume that $v \approx v_p$. To avoid confusion, EUHFORIA quantities will hereafter be denoted without indexes, while proton quantities will be denoted with index $p$. This approximation is of course different from known observations, but it would require to move from ideal MHD to at least two-fluid resistive MHD to overcome it, which is beyond the scope of this paper \citep{Priest2014}.

\subsection{CME modeling}
\label{subsec:cme_modeling}

EUHFORIA offers the possibility to inject a coronal mass ejection (CME) at 0.1~au inside the heliospheric part in a time-dependent way, in order to model its propagation and interaction with the modeled ambient solar wind. There are various models available to represent the CME. In this study, we will be using two of them.

First, we use the cone model, described in \cite{Pomoell2018} and similar to \cite{Odstrcil1999}. The CME is treated as a hydrodynamic cloud and is characterized by a constant angular width (which then defines an initial radius at the injection point), propagation direction and radial speed. The user can also prescribe the injection point, defined via its coordinates in HEEQ frame. The cross-section is assumed to be circular, and the CME density, pressure and radial speed are constant.

The other model we use is the linear-force-free spheromak (LFFS) model, described in \cite{Verbeke2019} and similar to \cite{Kataoka2009, Shiota2016}. It is a magnetized model, but unlike the model from \cite{Gibson1998}, the CME completely goes through the boundary without footpoints left attached to the solar surface. The CME, considered to be a sphere of radius $r_0$ upon the time of its injection, is launched outward. The velocity of the CME is chosen to be constant everywhere within the CME, and always oriented along the given propagation direction. As a result, the total velocity vector $v$ is not purely radial at the inner simulation boundary, but contains also a latitudinal and longitudinal component as well. This means that the total speed of the CME $v_{3D}$ can be decomposed into two components: the radial speed $v_{rad}$ and the expansion speed $v_{exp}$. In our case, we prescribe only the radial speed at the injection point. It requires the same physical input parameters as the cone model, plus three additional magnetic parameters: the handedness $H$ of the spheromak (which determines the polarity of the magnetic field inside the CME);
the tilt angle $\tau$ (measured from the $z$-axis in the $yz$-plane); the total toroidal flux $F$ (related to the magnetic field strength $B_0$). The spheromak orientation is thus given by the injection point, combined with the tilt that gives the axis of symmetry of the spherical structure (the magnetic configuration is shown in appendix in Figure~\ref{fig:handedness}). 
For example, in HEEQ coordinates of EUHFORIA, if the tilt is equal to 0, then the axis of symmetry is the vertical $z$-axis; if the tilt is equal to 90\degree, the axis of symmetry is the horizontal $y$-axis. Compared to Euler angles, the tilt angle corresponds to the elevation angle.
In the current implementation, we do not need additional angles (like the heading or bank angles) that would allow us to rotate around the axis of symmetry, since the spheromak exhibits symmetry in the azimuthal direction $\phi$.
The direction of propagation of the spheromak is finally set to be perpendicular to the 0.1~au boundary surface at the injection point.

\begin{figure}[th!]
    \centering
    \plotone{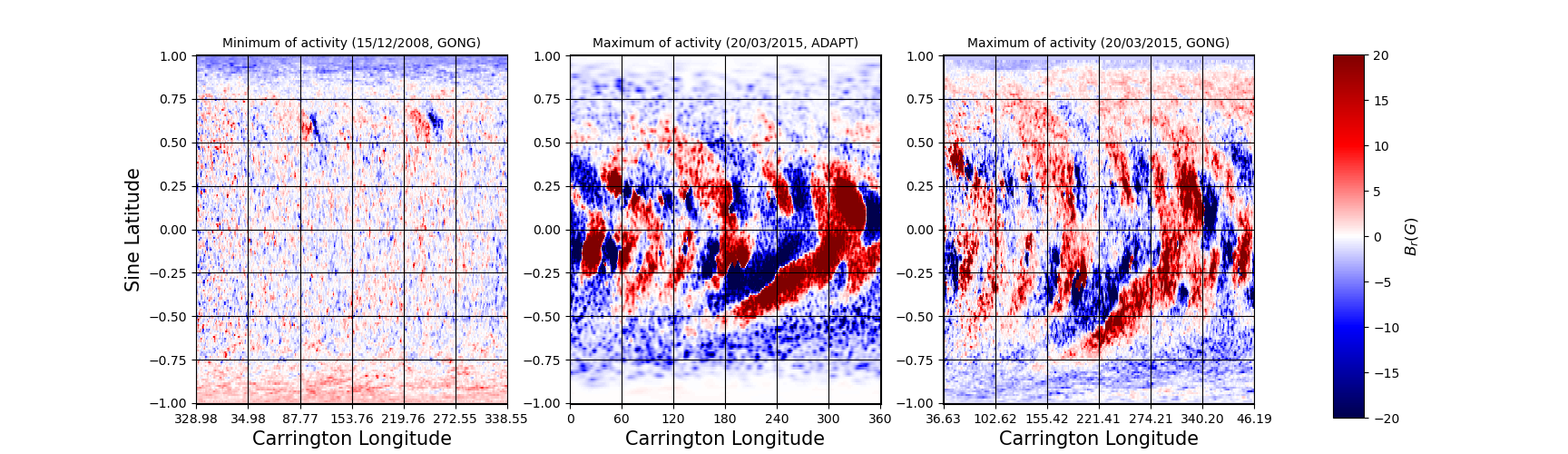}
    \caption{
    Comparison between the two selected magnetic maps of the radial magnetic field $B_r$ for minimum (left panel) and maximum (middle panel) of activity. The left panel shows a GONG synoptic map corresponding to the 15th of December 2008, representative of a minimum of activity (dipolar field with very few active regions). The middle panel shows a GONG-ADAPT map corresponding to the 20th of March 2015, representative of a maximum of activity (multipolar field with intense active regions). 
    We also show for information the GONG map corresponding to the date chosen for maximum of activity (right panel). We can thus see that there is more difference between minimum and maximum of activity than between GONG and GONG-ADAPT.
    Positive polarity of the magnetic field is shown in red, negative polarity in blue. The color bar has been set symmetric and saturated to better visualize the structures (between -20 and 20 G). Note that the y-axis for the left panel is in sine latitude.} 
    \label{fig:comp_bc_mag}
\end{figure}

\section{Boundary conditions} 
\label{sec:bc}

In this section, we discuss how we selected the boundary conditions at 0.1~au for our heliospheric study. We first explain in subsection \ref{subsec:bc_wind} the parameters related to the choice of the solar wind background, and then in subsection \ref{subsec:bc_cme} the parameters related to the ICME initialization.

\subsection{Solar wind for minimum and maximum of activity}
\label{subsec:bc_wind}

The idea behind this study is to use a solar wind background that would be more realistic than in previous studies using EUHFORIA: for example, \cite{Scolini2021_sir} used an ideal analytical background, and \cite{Asvestari2022} even approximated the magnetic field as a monopole (which is an impossible configuration due to the Maxwell equations). We especially aim to reflect the complexity brought by the 11-year modulation of the solar cycle. To do so, we first selected two dates that correspond to typical minimum and maximum of solar activity. 

For the minimum, we selected the Carrington Rotation number 2077, set between November 20 and December 17, 2008. This corresponds to the end of solar cycle 23. This date is indeed a well-known case used for calibration of coronal models at minimum of activity due to its very quiet magnetic field during this period \citep{Rusin2010, vanderHolst2014, Wiegelmann2017, Perri2022a}. It has even been chosen as the International Space Weather Action Team (ISWAT) validation benchmark for solar-wind models \citep{Reiss2022}. At this date, only the GONG magnetograms are available; the corresponding magnetogram can be found in the left panel of Figure~\ref{fig:comp_bc_mag}. It exhibits typical features from minimum of activity, such as strong flux concentration at the poles with different polarities in the two hemispheres, which is a marker of the dipolar-dominant structure of the solar magnetic field at this period. There are also very few active regions, so that we mostly see the super granulation pattern typical of the quiet Sun. 

\begin{figure}[th!]
    \centering
    \plottwo{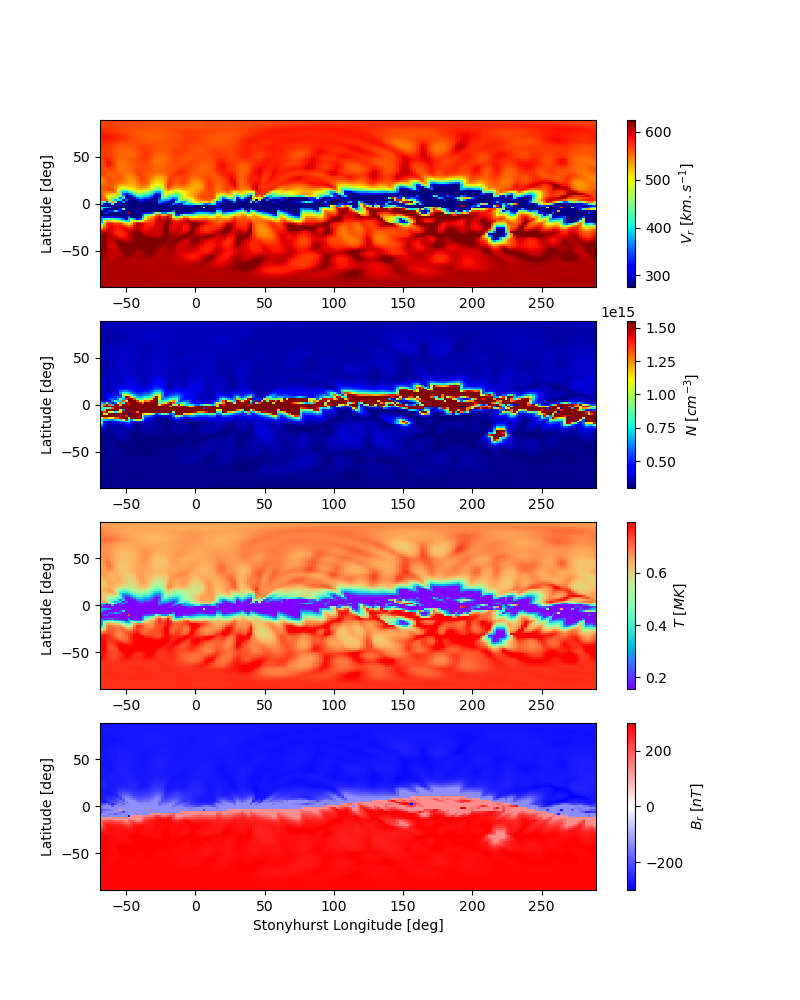}{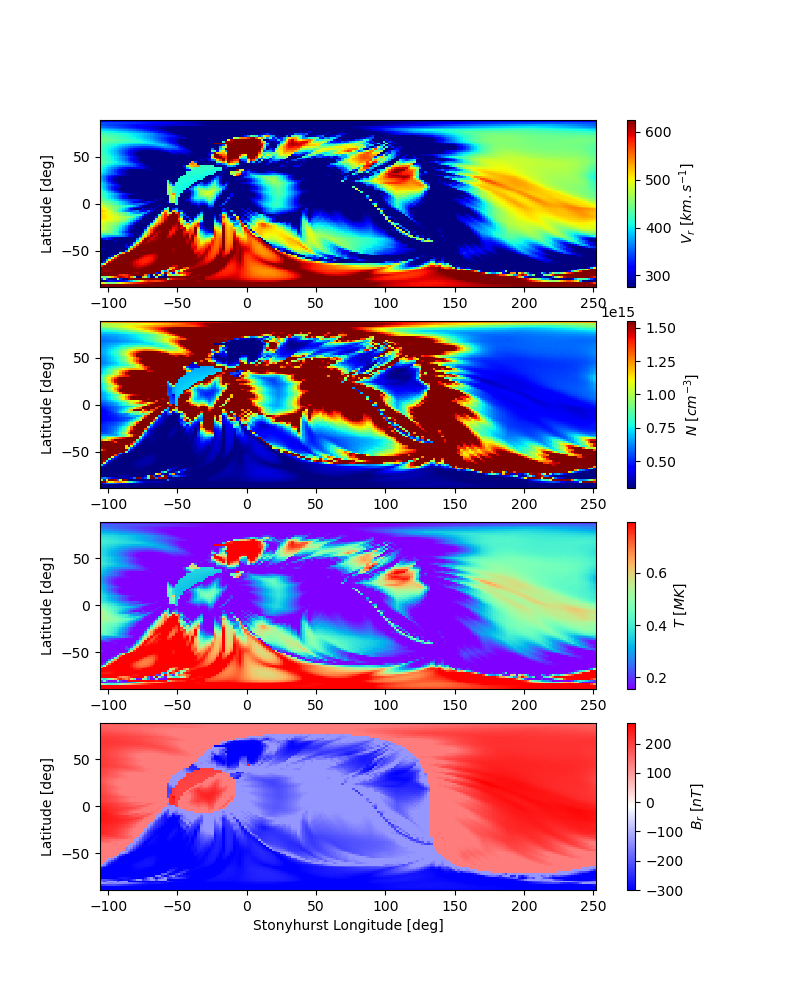}
    \caption{Comparison of the EUHFORIA input boundary condition at 0.1~au for minimum (left panel) and maximum (right panel) of activity. The first row shows the radial velocity $v_r$ in km/s, the second row the number density $N$ in $\rm{cm}^{-3}$, the third row the temperature $T$ in $10^6 \ K$, the fourth row the radial magnetic field $B_r$ in~nT. All these profiles were derived from the magnetic maps displayed in Figure~\ref{fig:comp_bc_mag} using Wang-Sheeley-Arge and PFSS+SCS empirical relations. X-axis is in Stonyhurst longitude, y-axis in latitude.}
    \label{fig:comp_bc_wsa}
\end{figure}

For maximum of activity, we selected the date of March 20, 2015. This date corresponds to the maximum of solar cycle 24, and has been extensively used as a benchmark date for many coronal models \citep{Yeates2018}. After 2010, the GONG-ADAPT maps are available, so we use it for the maximum as it reduces the probability of having spurious features at the solar poles thanks to the additional post-processing of ADAPT. This causes a difference in the provider and thus the processing of the input map. However, it guarantees that the final solution does not suffer from numerical artifacts. We also show in Figure \ref{fig:comp_bc_mag} the difference between GONG and GONG-ADAPT at maximum of activity. We can see that there is a difference in amplitude for the most intense active region, sometimes a difference in polarity for the quiet Sun and a different method for filling the solar poles. However, the general structure is still very similar, and these differences are still less than the difference between minimum and maximum of activity. This means that the differences we may see in the final CME solutions are indeed mainly due to the difference in solar activity. We will dedicate a specific study on the impact of the input magnetic map on the final CME solution, but for now it is out of the scope of this paper. The map can be seen in the right panel of Figure~\ref{fig:comp_bc_mag}. At maximum of activity, many active regions are present at the surface of the Sun between -0.5 and 0.5 in sine latitude (which corresponds to around -40 and 40$\degree$ in latitude), creating intense magnetic field configurations that dominate over the poles and the quiet Sun. 

The implication of these two different magnetic maps for the coronal part of EUHFORIA can be seen in Figure~\ref{fig:comp_bc_wsa}. We represent the boundary condition computed by the semi-empirical coronal model which will serve as boundary condition (BC) for the heliospheric MHD model. The left panel shows the BC for the selected minimum of activity, the right panel shows the BC for the selected maximum of activity. In each panel, the first row corresponds to the radial velocity in km/s, the second to the number density in $\rm{cm}^{-3}$, the third to the temperature in $10^6 \ K$ and the last one to the radial magnetic field component in nT. At minimum of activity (left panel), the corona is very structured, with fast hot wind at the poles and slow dense cold wind at the equator. The magnetic field is very much dipolar, with a negative polarity at the northern pole and positive polarity at the southern pole. The HCS between the two polarity is almost in the ecliptic with few disturbances. At maximum of activity, slow and fast wind are much more mixed at all latitudes. The HCS exhibits a more complex pattern, with the negative polarity from the southern hemisphere almost filling completely the map between longitudes 0 and 130. There is also an incursion of positive polarity in the negative polarity around longitude -50. These different BCs will result in different backgrounds for propagation of the CMEs, as will be detailed in Section \ref{subsec:hydro_wind}. Note that no optimization was performed here to constrain the wind with observations: first, because we do not aim at reproducing a specific event, this is foremost a theoretical study; second, because we want to keep an operational set-up to evaluate the implication for the space-weather forecasts done using EUHFORIA that use these automatic parameters.

\begin{table}[]
    \centering
    \begin{tabular}{|c||c|c|c|}
        \hline
        Case & Hydro CME & Reference CME & Median CME \\ \hline
        Type of CME model & Cone & LFFS & LFFS \\ \hline
        Latitude of center [deg. HEEQ] & 0.0 & 0.0 & 0.0 \\ \hline
        Longitude of center [deg. HEEQ] & 0.0 & 0.0 & 0.0 \\ \hline
        Radius [$R_\odot$] & 15.0 & 15.0 & 15.0 \\ \hline
        Speed [km/s] & 1400 & 763 & 541 \\ \hline
        Mass density [kg/m3] & 1.0 $10^{-18}$ & $1.0 \, 10^{-18}$ & $2.0 \, 10^{-18}$ \\ \hline
        Temperature [K] & $8.0 \, 10^5$ & $2.4 \, 10^5$ & $6.2 \, 10^5$ \\ \hline
        Handedness & N/A & $\pm$1 & $\pm$1 \\ \hline
        Tilt [deg.] & N/A & 0.0 & 0.0 \\ \hline
        Flux [Wb] & N/A & $1.0 \, 10^{14}$ & $2.3 \, 10^{13}$ \\ \hline
    \end{tabular}
    \caption{Summary of the CME parameters used at 0.1~au as a boundary condition for the EUHFORIA runs. For each case, we specify the model used, the center of the injected CME with its latitude and longitude in HEEQ coordinate, the radius of the inserted CME, the speed, mass density, and temperature. For the magnetized CMEs, we also specify the handedness, tilt and magnetic flux.}
    \label{tab:cme_params}
\end{table}

\subsection{Definition of the CME parameters}
\label{subsec:bc_cme}

At 0.1~au, we also have to define the parameters of the CME to inject in the simulation. In this study, we have chosen three cases to explore. All the corresponding parameters can be found summarized in Table~\ref{tab:cme_params}.

The first case is based on a real event that took place on July 12, 2012. This case has been extensively studied in \cite{Scolini2019} and \cite{Scolini2021_radial} using a magnetized LFFS model. The parameters for this CME were derived from observations, using white-light coronagraph images to constrain the geometrical and kinetic parameters, and photospheric and low-corona observations of the source active region to constrain the magnetic parameters \citep[for more details, see][]{Scolini2019}. The resulting parameters are summarized in column labeled Reference CME of Table~\ref{tab:cme_params}. They have been adapted in the scope of this study: we have adjusted the center of the CME to be along the Sun-Earth axis (which yields 0$\degree$ in HEEQ latitude and longitude), and we have selected the same average initial radius at injection for all cases of $15 \, R_\odot$ (to avoid geometric factors contributing to the variation in mass). This value corresponds to an average of the values found by GCS fitting for the events studied in \cite{Scolini2019}, which were 10.5, 14.5, 16.8 and 18.0 \p{$R_\odot$}. 
The injection point being on the Sun-Earth line in HEEQ coordinates, the CME direction of propagation is towards Earth, to simulate optimal initial conditions for a full hit.
The tilt has also been set to 0 to study a CME where the full magnetic ejecta propagates along the ecliptic plane. The speed, mass density, temperature, handedness and flux are however the same as in \cite{Scolini2019}. 

We also wanted to try a limit case with the same CME, but using the cone model instead of the LFFS model. This means that the CME is purely hydrodynamic (without an inner magnetic field). To obtain the corresponding parameters, we used this time the DONKI database\footnote{\url{https://kauai.ccmc.gsfc.nasa.gov/DONKI/}} which recommended a speed around 1400 km/s. The lower initial speed is due to the fact that we prescribe only the input radial speed, which is equal to the full 3D speed inferred from observations for a cone model ($v_{rad} = v_{3D}$), but equal to the difference between the 3D speed and the expansion speed for a spheromak ($v_{rad} = v_{3D} - v_{exp}$). For the reference case, the CME geometric parameters were derived using a GCS model (Graduated Cylindrical Shell model, see \cite{Thernisien2009, Thernisien2011}), which gave a full 3D speed $v_{3D}$ of 1266 or 1352 km/s depending on the fitting. The expansion speed of the CME is derived using empirical relations from \cite{DalLago2003} and \cite{Schwenn2005}, which is an alternative when 3D reconstruction of the event is not possible due to a single spacecraft configuration for the observation (more details in \cite{Scolini2019}). These relations give: $v_{rad} = 0.43 v_{3D}$ in the case of a spheromak, which explains the difference in input speed of a factor close to 2. The DONKI database values are thus consistent with the values obtained from \cite{Scolini2019}. This is explained in more details in \cite{Scolini2019}. We then completed the input values with the default EUHFORIA mass density and temperature (respectively $1.0 \, 10^{-18}$ kg/m3 and $8.0 \, 10^5$ K, as specified in \cite{Pomoell2018}). All other geometric parameters were kept identical to limit the number of free parameters of the study. This is summed up in the column labeled Hydro CME of Table~\ref{tab:cme_params}.

\begin{figure}[th]
    \centering
    \plotone{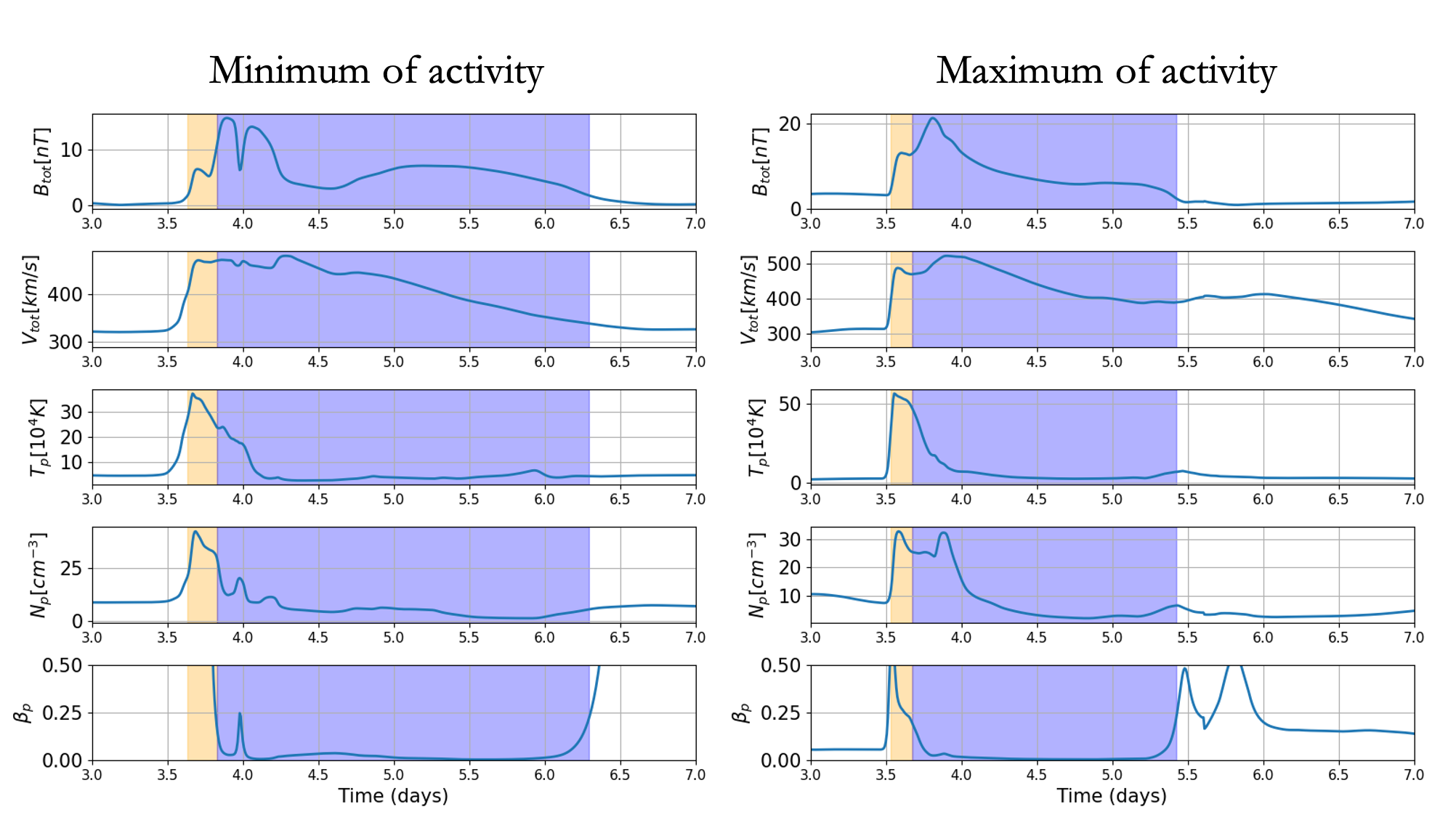}
    \caption{Comparison of the ICME parameters obtained at Earth for the median case at minimum (left panel) and maximum (right panel) of activity. The first row is the total magnetic amplitude in nT, the second row the total velocity amplitude in km/s, the third row the temperature in $10^4$ K, the fourth row the number density for the protons in $\rm{cm}^{-3}$ and the last row $\beta$ parameter of the plasma. A yellow rectangle highlights the region corresponding to the sheath, and a blue rectangle the region corresponding to the magnetic cloud. The method used to derive these regions is explained in more details in Appendix \ref{appendix:shock_sheath_mc}.}
    \label{fig:cme_median_profiles}
\end{figure}

Finally, we also wanted to have a magnetized case which would be more representative of an average CME at Earth. To obtain such values, we combined two studies. First, we used the statistical study from \cite{Regnault2020}, where 20 years of ACE data have been analyzed using the superposed epoch analysis (SEA) method to derive the distribution of ICME parameters at Earth. From this study, we can extract the median value of the distribution, which yields the following parameters:
\begin{equation}
    v_r \approx 450 \, \rm{km/s}, N \approx 5 \, \rm{cm}^{-3}, T \approx 4 \, 10^4 K, B_r \approx 10 \, \rm{nT}.
    \label{eq:florian_1au}
\end{equation}
By using the median, we select a set of parameters so that 50\% of ICMEs have values under this set, and 50\% above. We could have also used the mode (which is the most probable set of parameters), but due to the fact that it is a log normal distribution, it would have been biased towards slower events that would have been too close to our background wind speed to allow proper analysis. 
Once we have these values at 1~au, we need to extrapolate them back to 0.1~au. To do so, we use the scaling laws derived by \cite{Scolini2021_radial} by studying the radial evolution of ICMEs in EUHFORIA:
\begin{equation}
    v_r \propto r^{-0.08}, N_p \propto r^{-2.38}, T_p \propto r^{-1.19}, B_r \propto r^{-1.9}.
\end{equation}

Combining these two set of parameters, we can obtain the corresponding values at 0.1~au for a median CME:
\begin{equation}
    v_r = 541 \, \rm{km/s}, N_p = 1.2 \, 10^3 \, \rm{cm}^{-3}, T_p = 6.2 \, 10^5 \, K, B_r = 7.9 \, 10^2 \, \rm{nT}.
\end{equation}
We then convert these values to obtain the parameters needed for the input configuration file of EUHFORIA, using the following relations: 
\begin{equation}
    \rho = m_pN_p\times 10^6, \phi_t = \frac{2B_0}{\alpha}\left[-\rm{sin}\left(\alpha r_0\right) + \int_{0}^{\alpha r_0}\frac{\rm{sin}\,t}{t}dt\right],
\end{equation}
with $m_p$ being the proton mass, $\alpha r_0=4.4934$ (which corresponds to prescribing the radial component of the magnetic field to be 0 on the boundary of the spheromak) and $r_0=15 R_\odot$ (for more details, see \cite{Verbeke2019}).
The resulting parameters are summed up in the column labeled Median CME of Table~\ref{tab:cme_params}. Compared to the real event, we can see that the median CME is slower, denser and hotter, and less magnetized than the reference case.

We finally check with EUHFORIA simulations that this approach yields correct results by analyzing the values of the ICME obtained at Earth for the median CME, both at minimum and maximum of activity. Results can be visualized in Figure~\ref{fig:cme_median_profiles}. The minimum of activity case in on the left panel, the maximum of activity case on the right panel. For each case, we plot the total magnetic amplitude in nT, the total velocity amplitude in km/s, the temperature in $10^4$ K, the number density for the protons in $\rm{cm}^{-3}$ and the $\beta$ parameter of the plasma, defined as $\beta = P_{th}/P_{mag} = 2 P_{p,th}/P_{mag} = 2\beta_p$ (thermal pressure over magnetic pressure). The panel is made to be easily comparable with the analysis from \cite{Regnault2020}, which has the same layout for the SEA. On top of the curves, we plot a yellow rectangle to highlight the region corresponding to the sheath and a blue rectangle for the region corresponding to the magnetic ejecta. The method used to derive these regions is explained in more details in Appendix \ref{appendix:shock_sheath_mc}. In the magnetic ejecta, we obtain on average the following values for the minimum of activity case: 
$v_r \approx 420$ km/s, $N \approx 5 \, \rm{cm}^{-3}$, $T \approx 5 \, 10^4$ K, $B \approx 8$~nT. Similarly, we obtain, for the maximum of activity case: $v_r \approx 450$ km/s, $N \approx 10 \, \rm{cm}^{-3}$, $T \approx 4 \, 10^4$ K, $B \approx 10$~nT. These values are consistent with the targeted valued at 1~au (cf. equation (\ref{eq:florian_1au})), which means these cases can indeed be classified as representative of a median ICME as seen at Earth. 
 
\begin{figure}[th]
    \centering
    \plotone{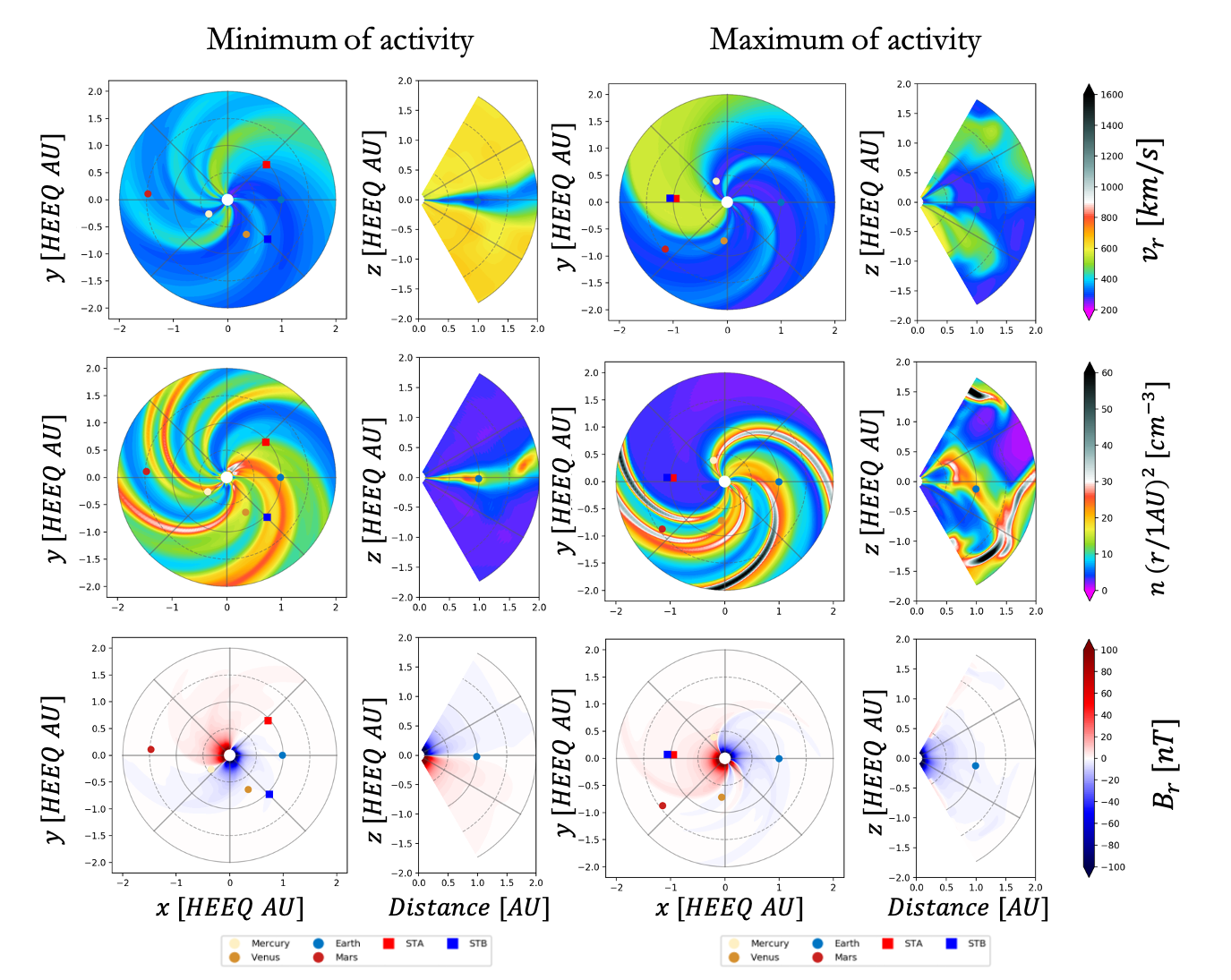}
    \caption{Comparison of the heliospheric wind background at minimum (left panel) and maximum (right panel) of activity. The first row is the radial wind speed in km/s, the second row the number density in $\rm{cm}^{-3}$ normalized to 1~au and the last row the radial magnetic field in nT. For each row, we show the ecliptic (view from above) and meridional (view from the side crossing Earth) views in the HEEQ frame. Various planets and satellites are also shown in the bottom legend.
    }
    \label{fig:comp_bc_wind_2d}
\end{figure}

\begin{figure}[th]
    \centering
    \plotone{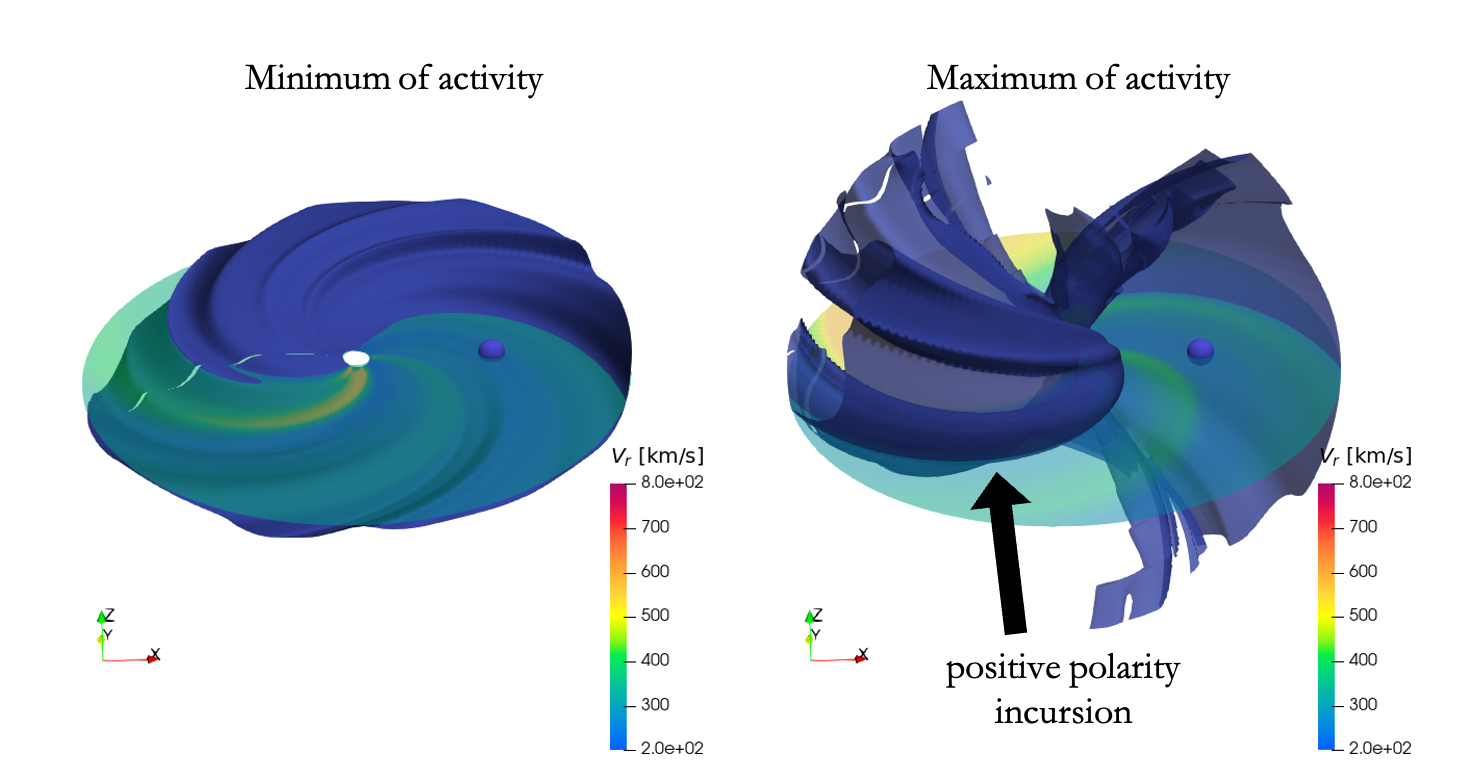}
    \caption{Comparison of the 3D heliospheric current sheet (HCS) structure for the selected minimum (left panel) and maximum of activity (right panel). The HCS contour is shown in dark blue. For context, the ecliptic plane with the radial velocity is shown in transparency. 
    The position of the Earth is indicated by a blue sphere. Both figures are in HEEQ frame.}
    \label{fig:comp_wind_hcs_3d}
\end{figure}

\begin{figure}[th]
    \centering
    \plotone{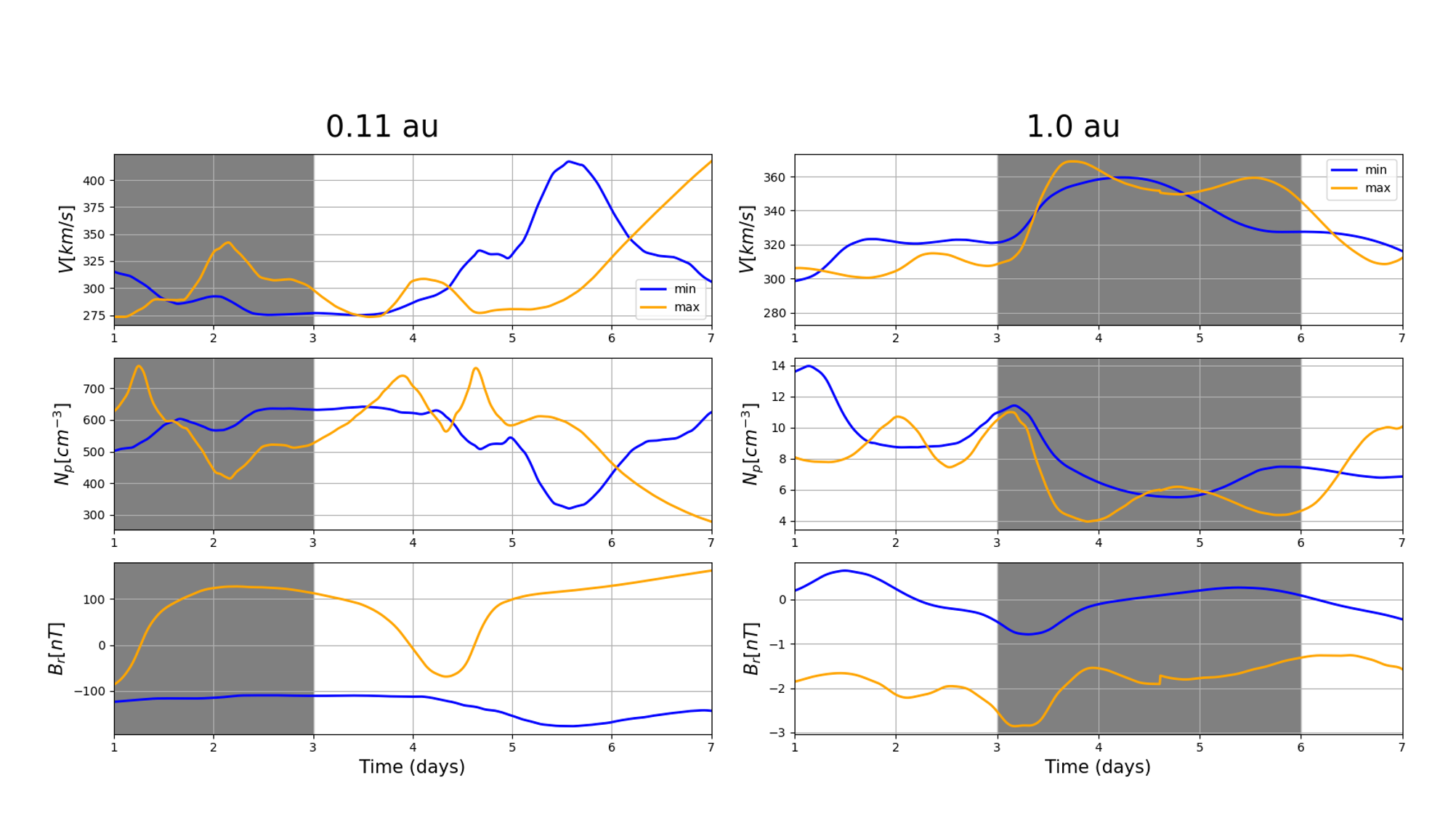}
    \caption{Comparison of the time evolution of the background solar wind parameters at 0.11~au (left panel) and 1~au (right panel) along the Sun-Earth axis. The first row shows the total velocity in km/s, the second row the number density in $\rm{cm}^{-3}$ and the last row the radial magnetic field component in nT. The curves for the minimum of activity are in blue, while the curves for the maximum of activity are in orange. Grey rectangles have been added to show the time intervals when the background solar wind is likely to interact with the ICME.}
    \label{fig:comp_wind_1d}
\end{figure}

\section{Limit cases}
\label{sec:limit_cases}

Before analyzing further the differences between the ICMEs, we want to analyze two limit cases to better understand the context of our simulations. In Section \ref{subsec:hydro_wind}, we will first present a case with only the background wind (no CME inserted) to better quantify the differences and similarities between without and with CME. 
In Section \ref{subsec:hydro_cme}, we will then inject a hydrodynamic CME (described in column labeled Hydro CME of Table~\ref{tab:cme_params}) to quantify the impact of the hydrodynamic parameters alone.

\subsection{Wind-only simulation}
\label{subsec:hydro_wind}

In this first limit case, we do not inject any CME, we just let the wind background evolve for 7 days to see how it behaves without perturbation. In Figure~\ref{fig:comp_bc_wind_2d}, we can see the heliospheric part of EUHFORIA with the modeling of the ambient solar wind background. We show the radial wind speed, the number density and the radial magnetic field, both in the ecliptic and meridional (including Earth) planes.  
This allows us to get a first qualitative look at the similarities and differences between the selected minimum and maximum of activity. Since these are realistic solar wind backgrounds, we can see a lot of substructures in the wind. One constrain on the dates is that any large-scale structures (such as high-speed streams where the wind speed reaches 500 km/s or more) is not Earth directed and thus not geo-effective, to limit their interference with the propagation of the CME towards Earth. The dates have been adjusted so that the sector in the ecliptic plane is dominantly negative (in blue in the bottom row) in both cases. 

There are however expected differences that are relevant to this study, because they are representative of what we expect from minimum and maximum of activity configurations. At minimum of activity, the meridional view shows a very organized wind structure, with slow dense wind in the ecliptic plane, and fast less dense wind near the poles. This is a logical consequence of the structures observed in the coronal boundary condition in Section \ref{subsec:bc_wind}. At maximum of activity, the solar wind organization is much more complex. For the magnetic field, we also see that at minimum of activity, the current sheet location near the ecliptic results in the polarity sectors being very clearly defined: at Earth location, the northern hemisphere is negative while the southern hemisphere is positive. At maximum of activity, the polarity remains negative in latitude, but a positive sector will cross the Earth later on in the simulation (see the bottom right panel of Figure~\ref{fig:comp_bc_wind_2d}). This kind of fast polarity switch is expected at maximum of activity and is one of the features we are interested in.

To focus on the magnetic field structure, in Figure~\ref{fig:comp_wind_hcs_3d}, we represent the HCS in 3D. We can clearly see its quasi-ecliptic shape at minimum of activity, with the Earth being positioned slightly above it. At maximum of activity and in the inertial frame, the current sheet is very distorted, so that the Earth falls into one sector. The contour on the left-side of the picture is the positive polarity incursion from the ecliptic view, and it may interfere with the CME propagation during its latest stages.

Finally, in Figure~\ref{fig:comp_wind_1d}, we show a more quantitative view of the background solar wind with 1D evolution plots. We plot the same quantities as in Figure~\ref{fig:comp_bc_wind_2d}, but along the Sun-Earth axis using virtual satellites. The left panel at 0.11~au gives a view of what the background looks like at the injection point of the CME, while the right panel shows the background at 1~au to show the ambient medium at Earth. Minimum and maximum of activity data are over-imposed, respectively in blue and orange. Grey rectangles have been added to highlight the regions of interest in each panel for the interaction with the ICME. We notice in the right panel that the wind structures are indeed very similar at Earth, except for $|B_r|$ which is more important at maximum of activity (as expected since the solar activity is increasing). Close to injection point (which means on the left panel), the wind speed is low in both cases (around 300 km/s). The main difference lies in the polarity inversion: at minimum of activity, the magnetic field polarity remains mostly negative, getting weak around 1~au (Figure~\ref{fig:comp_wind_1d} right panel), while at maximum of activity the disturbed HCS allows for a polarity switch from negative to positive as the ICME is progressing outward.

\begin{figure}[th]
    \centering
    \plotone{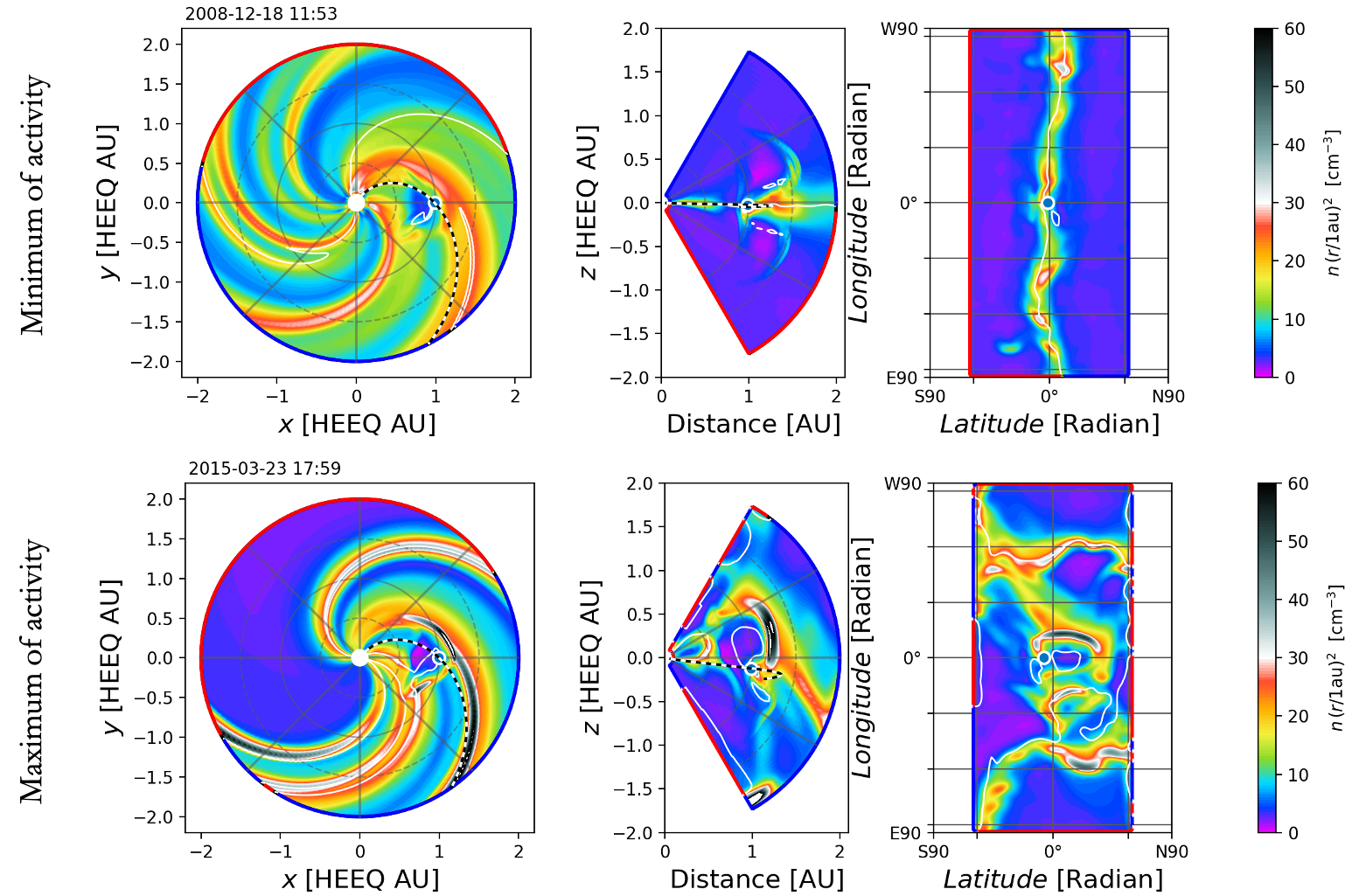}
    \caption{ 
    Comparison of the cone CME propagation for the minimum (top panel) and maximum (bottom panel) of activity cases. We show the density in $\rm{cm}^{-3}$, normalized to 1~au to better visualize the CME ejecta as an under-dense structure. We show the moment when the ejecta is reaching Earth (symbolized by a blue circle circled by a white line to the right of the Sun). For each case, we show the ecliptic (view from above, left panel), meridional (view from the side including Earth, middle panel) and spherical (view at 1~au, right panel) views in the HEEQ frame. The magnetic field line connecting the Earth to the Sun is shown in dotted line. The polarity of the magnetic sectors is shown at the edges of the frames for reference (red for positive, blue for negative), separated by the HCS shown as a white line inside the domain. Animated versions of this figure are available in the online version, showing the full propagation of the cone CME from 0.1~au to the Earth for both the minimum and maximum of activity backgrounds, with the evolution of the radial magnetic field, density and radial velocity.}
    \label{fig:cone_2d_comp}
\end{figure}

\begin{figure}[th]
    \centering
    \plotone{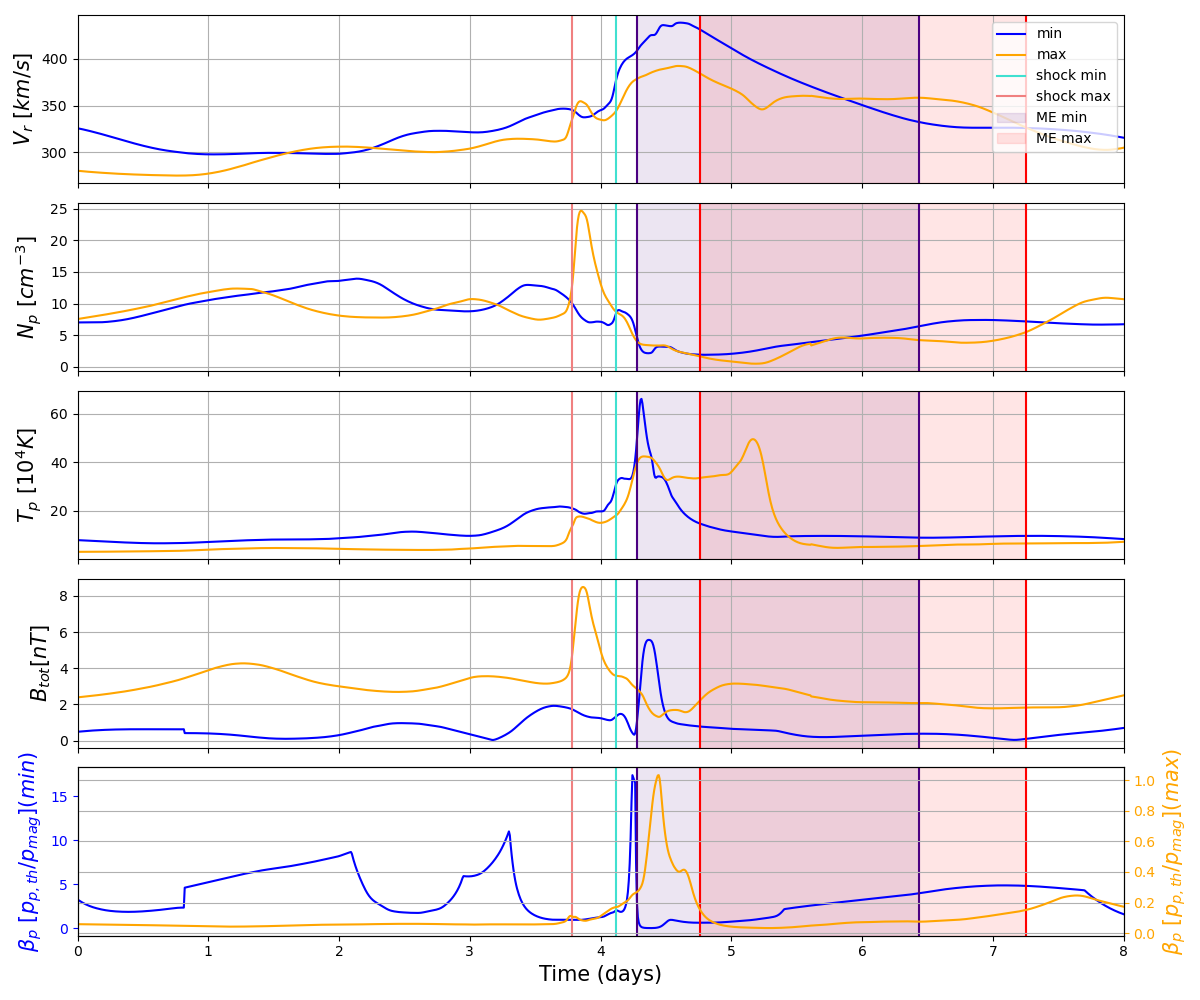}
    \caption{Comparison of the time evolution of the main physical quantities at 1~au at Earth position for the propagation of the cone CME. The time is the number of physical days that lasted the simulation. The blue line shows the minimum of activity case, while the orange line shows the maximum of activity case. From top to bottom, we show the radial velocity (in km/s), the proton number density (in $\rm{cm}^{-3}$), the temperature (in units of $10^4$ K), the total magnetic field (in nT) and the plasma beta parameter. The arrival of the shock is shown by a vertical line (light blue for minimum of activity case, red for maximum of activity case). The crossing of the ejecta is shown by a colored window (light purple for minimum of activity, light red for maximum of activity).
    }
    \label{fig:cone_1d_comp}
\end{figure}

\subsection{Hydrodynamic cone CME}
\label{subsec:hydro_cme}

In this second limit case, we will now inject a CME, but a purely hydrodynamic one. To do so, we use the cone model, detailed in Section \ref{subsec:cme_modeling}, and with the parameters described in the column labeled Hydro CME of Table~\ref{tab:cme_params}. As explained in Section \ref{subsec:bc_cme}, this corresponds to the typical parameters prescribed to study the event of the 2nd of July 2012, which is our reference case due to the great number of studies using EUHFORIA on it. This was a single-CME event that triggered a strong magnetic storm on Earth, which also explains why it is relevant for our study. The parameters have been adjusted to make the CME fully directed towards Earth in order to maximize the geo-effectiveness of the event. We inject the same cone CME in the two wind backgrounds corresponding to minimum and maximum of activity. This allows us to quantify the impact of the background solar wind, in a context where there is no magnetic interaction possible between the CME and the background (because the CME does not possess an intrinsic magnetic structure).

The results are displayed in Figure~\ref{fig:cone_2d_comp}. We show the density in $\rm{cm}^{-3}$ to better visualize the CME ejecta as an under-dense structure. We show the moment when the ejecta is reaching Earth (symbolized by a blue circle to the right of the Sun). For each case, we show the ecliptic (view from above), meridional (view from the side) and spherical (view at 1~au) views in the HEEQ frame. From the ecliptic view, we can see that the CME structures seem rather similar, although the shock for the maximum of activity case is better defined. The difference is more visible in the meridional view. 

For the minimum of activity case, the CME is actually cut in half, with a dragging middle structure and two accelerated lobes over and under the current sheet. This can easily be understood if we put this result in perspective with the wind structure at minimum of activity described in the previous section: since at minimum of activity the heliosphere is very organized, the parts of the CME caught in the current sheet are slowed down by the equatorial slow wind, while the parts outside the current sheet are accelerated by the fast solar wind. We can wonder how realistic it is to have a CME injected right in the middle of the current sheet at 0.1~au. Although CMEs have very little chances of being generated at the equator because they may face magnetic trapping \citep{Sahade2022}, they can be channeled towards the current sheet if they form close to the border of a streamer or pseudo-streamer \citep{Zuccarello2012}. Also, this effect is dominant at minimum of activity thanks to the dipolar structure of the solar magnetic field, which justifies even more this result for the minimum case \citep{Lavraud2014}. Similar configurations have been found for simulations of stellar CMEs associated with extremely dipolar stars \citep[and hence very organized astrospheres, ][]{Alvarado-Gomez2022}. At maximum of activity, on the other hand, the CME is much more compact, but undergoes a deflection towards the northern hemisphere. 

In the spherical view, we can see more clearly the impact of the ICME at Earth. At minimum of activity, we see the full hit of the central part of the ejecta with an under-dense structure at the same longitude and latitude than Earth. At maximum of activity, detecting the ICME signature is more difficult because of the complex structure of the heliosphere. The large under-dense structure north to the Earth is the ICME, while the largest one in the southern hemisphere is the trace of the open southern coronal hole.

To be more quantitative, we show in Figure~\ref{fig:cone_1d_comp} the time evolution of the main physical quantities at the position of the Earth at 1~au. We do this by using a virtual satellite within the simulation. From top to bottom, we show the radial velocity (in km/s), the proton number density (in $\rm{cm}^{-3}$), the temperature (in units of $10^4$ K), the total magnetic field (in nT), and the plasma beta parameter. The blue line shows the minimum of activity case, while the orange line shows the maximum of activity case. The horizontal axis shows the number of physical days after the insertion of the CME for both cases. The arrival of the shock is shown by a vertical line (light blue for minimum of activity case, red for maximum of activity case). The crossing of the ejecta is shown by a colored window (light purple for minimum of activity, light red for maximum of activity). 
Although the two CMEs have very different 3D structures, their 1D profile at Earth are actually not so different in density, temperature, and especially radial velocity. Because it is a cone CME, the detection of the ICME and its internal borders is a bit more challenging due to the lack of internal magnetic structure, but the global overview of all these physical quantities allow us to make estimations. The initial shock of the ICME is visible in velocity, but more clear in density and in the total magnetic field. From the radial velocity, we see that the shock is rather gradual, starting around day 4. The CME at minimum of activity is a bit faster, going over 400 km/s in the sheath. This is probably due to the fact that it was a full hit, compared to the maximum of activity case which was a flank hit. 
We do show the magnetic field in this case, but we remind the reader that it is not representative of the polarity of the ICME itself, but rather of the accumulation of magnetic field from the solar wind background in front of it due to the propagation of the shock. The temperature is mostly shown for context of the variations of the $\beta_p$ parameter. The plasma beta parameter is shown as to visually support our settings of the boundaries of the ejecta for both cases.
Because of the clear shock in density and total magnetic field, we can clearly say that the ICME traveling at maximum of activity arrives faster (before 4 days, vs after 4 days for the minimum case). However, if we look at the $\beta$ parameter, we can see that the sheath region is shorter at minimum of activity (it passes Earth in only 3.83 hours), so that in the end the magnetic ejecta arrives faster at minimum than at maximum (4.28 days vs. 4.76 days). The ejecta duration is rather similar (a bit more than 2 days), so logically the ejecta ends sooner for the minimum of activity. All these results are summarized in Table \ref{tab:timings}.

Please note that these results are specific to the background chosen, so it is not clear how generalized they can be. It can also be slightly affected by our method selected to detect the ICME borders (see Appendix \ref{appendix:shock_sheath_mc} for more details).

In conclusion, these two analysis show that for our purely hydrodynamic CME, the change in activity we used caused mostly a geometric deviation : our minimum case CME is decelerated in the ecliptic plane and accelerated out of it due to the heliosphere structure; our maximum case presents a deviation towards the northern hemisphere. In the 1D profiles, the ICME shock at maximum of activity seems to arrive faster, but the ME arrives later than in minimum. It also reinforces the limitation of one-vantage in situ measurement, since our two different CMEs actually produced similar speed 1D profiles at Earth which do not reflect the major geometrical difference in 3D.

\begin{figure}[th!]
    \centering
    \plotone{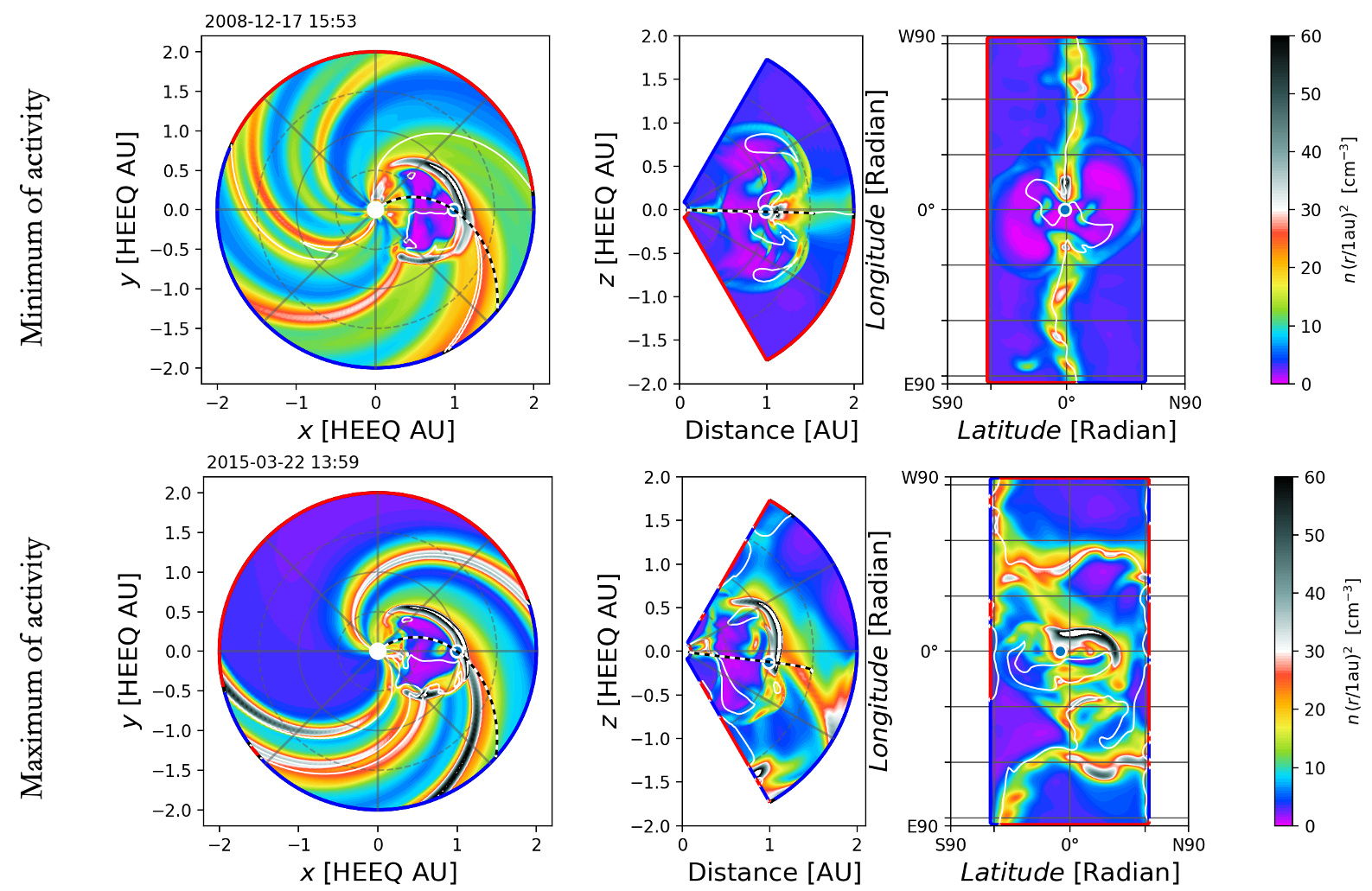}
    \caption{
    Comparison of the reference spheromak CME propagation for the minimum (top panel) and maximum (bottom panel) of activity cases. In this case, both CMEs have positive handedness. We show the density in $\rm{cm}^{-3}$, normalized to 1~au to better visualize the CME ejecta as an under-dense structure. We show the moment when the ejecta is reaching Earth (symbolized by a blue circle to the right of the Sun). For each case, we show the ecliptic (view from above, left panel), meridional (view from the side, middle panel) and spherical (view at 1~au, right panel) views in the HEEQ frame. The magnetic field line connecting the Earth to the Sun is shown in dotted line. The polarity of the magnetic sectors is shown at the edges of the frames for reference (red for positive, blue for negative), separated by the HCS shown as a white line inside the domain. Animated version of this figure are available in the online version, showing the full propagation of the reference spheromak CME from 0.1~au to the Earth for both the minimum and maximum of activity backgrounds, with the evolution of the radial magnetic field, density and radial velocity.
    }
    \label{fig:extreme_pos_2d_comp}
\end{figure}

\begin{figure}[th]
    \centering
    \plotone{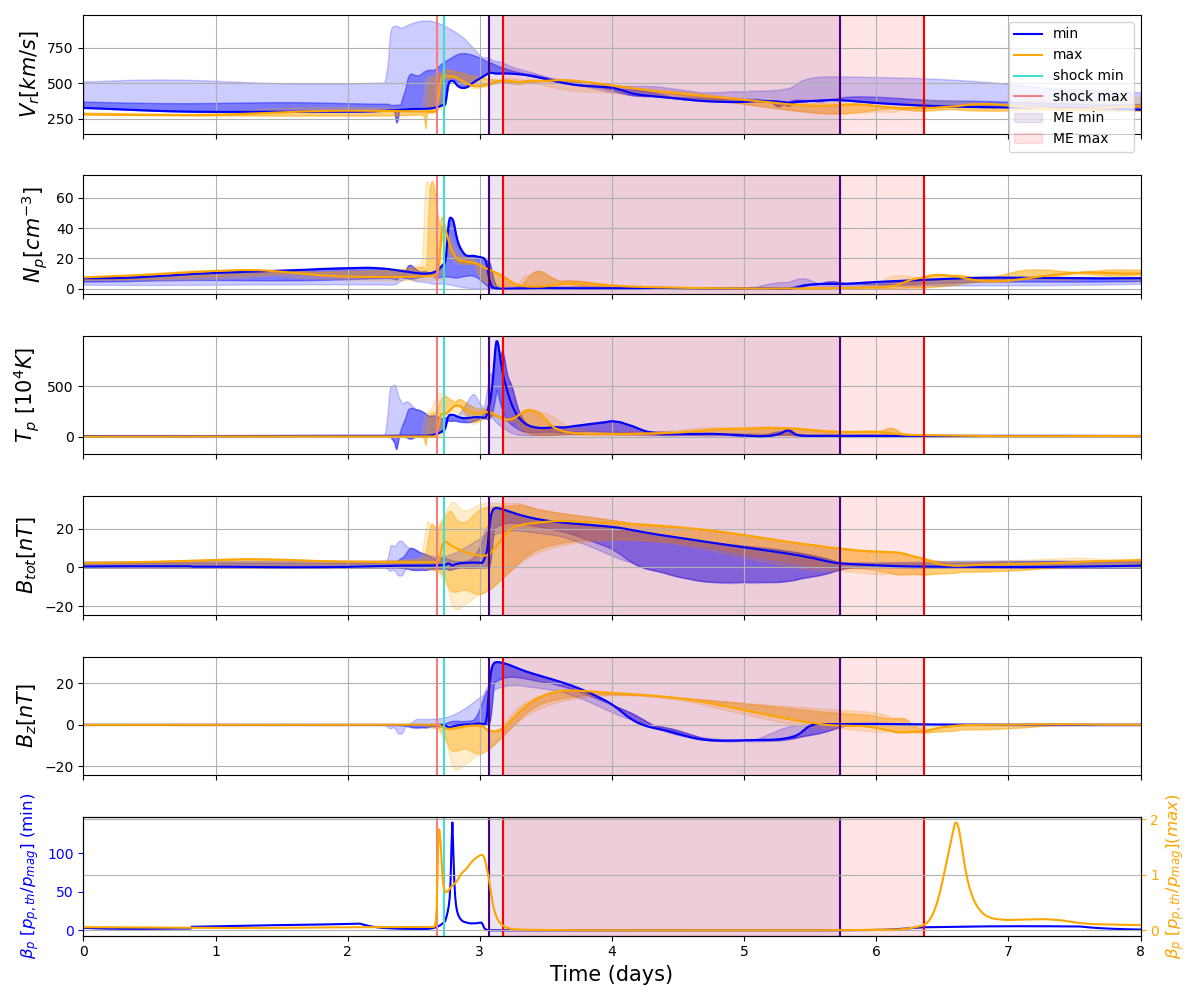}
    \caption{Comparison of the time evolution of the main physical quantities at 1~au at Earth position for the propagation of the reference spheromak CME with positive handedness. The time is 
    the number of physical days that lasted the simulation. The blue line shows the minimum of activity case, while the orange line shows the maximum of activity case. From top to bottom, we show the radial velocity (in km/s), the proton number density (in $\rm{cm}^{-3}$), the temperature (in units of $10^4$ K), the total magnetic field (in nT), the $z$ components of the magnetic field in HEEQ coordinates (in nT) and the plasma beta parameter. The arrival of the shock is shown by a vertical line (light blue for minimum of activity case, red for maximum of activity case). The crossing of the ejecta is shown by a colored window (light purple for minimum of activity, light red for maximum of activity). We also show the vertical deviation at 5 and 10 degrees north and south of the ecliptic plane as colored shaded areas around the curves.}
    \label{fig:extreme_pos_1d_comp}
\end{figure}

\section{Reference spheromak case from real event}
\label{sec:cme_12_july}

In this next section, we now use a different modeling for the CME with intrinsic magnetic field, which is the spheromak CME (described in Section \ref{subsec:cme_modeling}). With the previous section isolating the hydrodynamic effects, this allows us to better quantify the impact of the interaction of the CME magnetic field with the surrounding background. It becomes especially important that the magnetic field is more intense and more complex at maximum of activity. As explained in Section \ref{subsec:bc_cme}, we will first use parameters inspired by a true event from July 12 2012 which has already been extensively studied with EUHFORIA, and that will act as a reference case. The input parameters are summed up in Table~\ref{tab:cme_params}. With the internal magnetic field comes three new parameters: the tilt, flux intensity and handedness. 

In this first configuration, we inject a spheromak with positive handedness, which means the inner flux-rope is a right-handed helix at injection. In order to get a more compact study, negative handedness is only reported in the next section (Section \ref{sec:cme_median_neg}).  For more information about the handedness, please refer to appendix \ref{appendix:handedness}. The resulting CMEs can be seen in Figure~\ref{fig:extreme_pos_2d_comp}. We recover mainly similar effects observed in Section \ref{subsec:hydro_cme}: at minimum of activity the CME appears to be cut in half due to the highly-organized structure of the heliosphere, while at maximum of activity the CME is deflected to the northern hemisphere and thus producing a flank hit. When we compare it with Figure~\ref{fig:cone_2d_comp}, we can also see some differences. In both cases, the CME appears more structured and coherent, which is due to the inner magnetic field allowing for a more cohesive structure. This also produces a more defined shock at the front of the CME in both cases, especially visible in the density structure with a dark ring showing the corresponding over-density (the same color scale is used in both figures). At minimum of activity, the dislocation of the CME is less pronounced,  due to the inner magnetic field counteracting the solar wind influence. This is especially visible in the spherical view where this time the under-dense structure is visible not only at Earth latitude, but also north and south of it. At maximum of activity, the CME is more elongated in latitude and presents a slight asymmetry, probably due to the more complex magnetic interaction and reconnection with the irregular HCS magnetic configuration.

To be more quantitative, we also show the 1D evolution at Earth in Figure~\ref{fig:extreme_pos_1d_comp}, similar to Figure~\ref{fig:cone_1d_comp}. To complete this figure, we also show this time the vertical deviation at 5 and 10 degrees north and south of the ecliptic plane as colored shaded areas around the curves, similar to what was done in Figure 17 of \cite{Scolini2019}. This allows us to display some estimation of the uncertainty for the ICME profile around the Earth position.
We can see in this case that the shock is indeed more defined in all the physical quantities, arriving at Earth around 2.75 days. The ICME is thus arriving faster than in the pure hydro case, even though the initial speed of the CME is much less (but we recall from section \ref{subsec:bc_cme} that the full 3D speed is actually the same). 
Here the CME at maximum of activity still arrives first, but with only a 2-hour lead compared to the CME at minimum of activity, as seen more clearly in the radial velocity. The sheath is then followed by a magnetic ejecta clearly visible in the $B_z$ component, until the parameters return to the initial solar wind before the shock. Once again, the speed of the magnetic ejecta is slightly higher for the minimum of activity case, especially in the deviation at 5 and 10 degrees because of the acceleration of the polar solar wind. Density remains similar in both cases. The magnetic field component, this time representative of the structure of the inner flux-rope, is however different. 
At minimum of activity, $B_z$ rises first to 20~nT, before going to a -8~nT phase at the end of the ejecta. On the other hand, at maximum of activity $B_z$ always remain positive, but under 20~nT.
This difference in $B_z$ is very likely due to the difference in geometry discussed above, which results in a different impact parameter for both cases.
In this scenario, none of the CMEs are geo-effective, as their $B_z$ are both mostly positive.

This case shows that with an internal magnetic field, we retrieve similar results for the geometry of the propagation of the CMEs, except that the ejections are more cohesive thanks to their internal flux-rope. In this case, the CMEs arrive at a very similar time at Earth, with only a 2-hour lead in the maximum of activity case.

\section{Median spheromak case}
\label{sec:cme_median}

The previous results were interesting, but we can easily argue that they may hold only for the specific case we are studying. Although we do realize and acknowledge that our results are intrinsically dependent on the choice of our solar wind background and CME model, we include one last case that would aim at being slightly broader. To do so, we have defined in Section \ref{subsec:bc_cme} a median CME, based on the statistical analysis of ACE data from \cite{Regnault2020}, combined with extrapolations from 1~au to 0.1~au by \cite{Scolini2021_radial}. The parameters selected are detailed in Table~\ref{tab:cme_params}. We recall that compared to the previous case, the median CME is slower (541 vs.\ 763 km/s at injection), more dense (2 vs.\ 1 $10^{-18}$ $\rm{kg.m^{-3}}$ at injection), hotter (6.2 vs.\ 2.4 $10^5$ K at injection) and less magnetized (2.3 vs 10 $10^{13}$ Wb flux at injection). We will now perform the same study as in the previous section, but for this median CME, in order to see how general our results can be extrapolated.

\begin{figure}[th]
    \centering
    \plotone{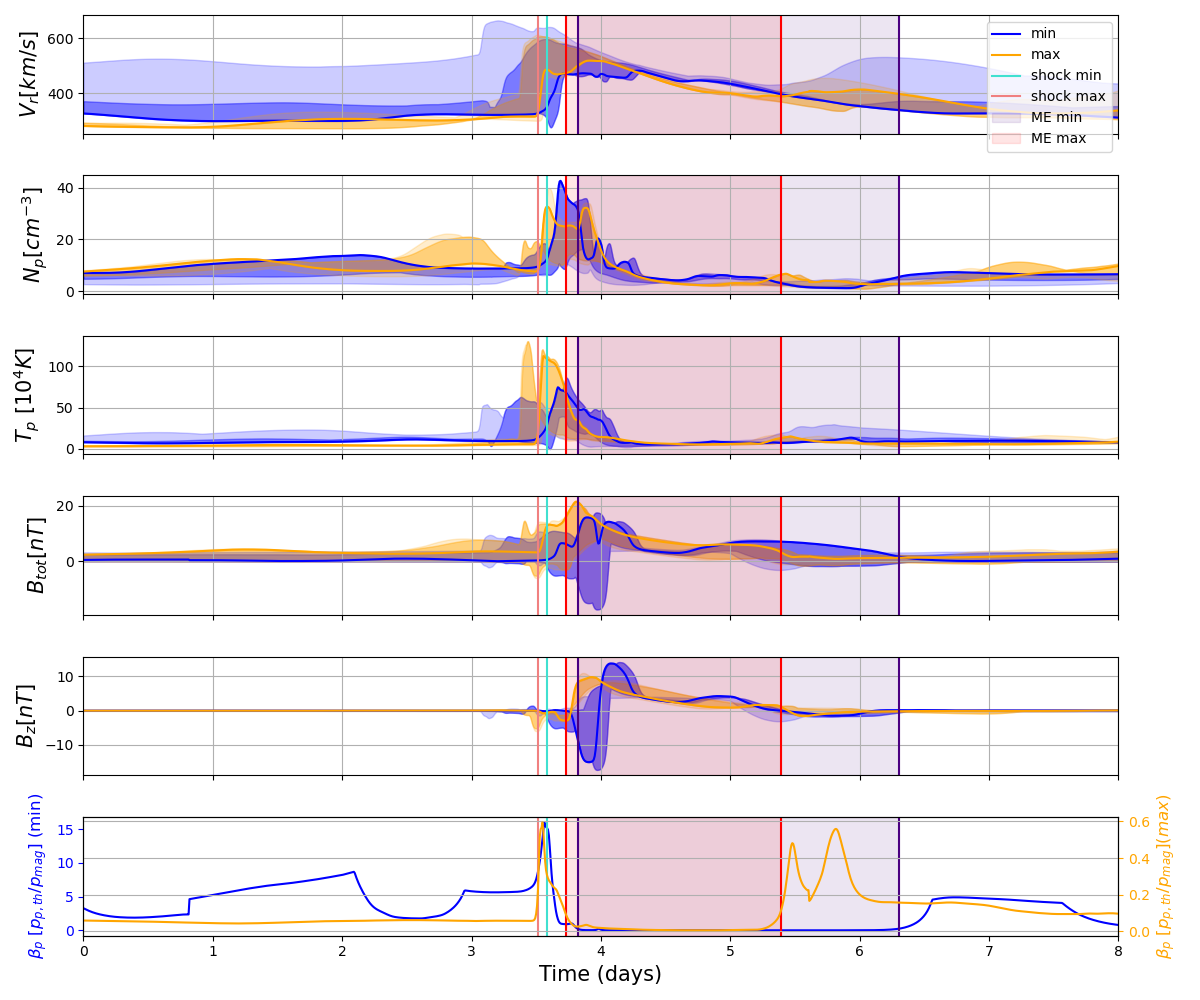}
    \caption{Comparison of the time evolution of the main physical quantities at 1~au at Earth position for the propagation of the median spheromak CME with positive handedness. The time is 
    the number of physical days that lasted the simulation. The blue line shows the minimum of activity case, while the orange line shows the maximum of activity case. From top to bottom, we show the radial velocity (in km/s), the proton number density (in $\rm{cm}^{-3}$), the temperature (in units of $10^4$ K), the total magnetic field (in nT), the $z$ component of the magnetic field in HEEQ coordinates (in nT) and the plasma beta parameter. The arrival of the shock is shown by a vertical line (light blue for minimum of activity case, red for maximum of activity case). The crossing of the ejecta is shown by a colored window (light purple for minimum of activity, light red for maximum of activity). We also show the vertical deviation at 5 and 10 degrees north and south of the ecliptic plane as colored shaded areas around the curves. Animated movies corresponding to this figure are available in the online version, showing the full propagation of the median spheromak CME with positive handedness in 2D cuts (equatorial, meridional and spherical) from 0.1~au to the Earth for both the minimum and maximum of activity backgrounds, with the evolution of the radial magnetic field, density and radial velocity.}
    \label{fig:median_pos_1d_comp}
\end{figure}

\subsection{Positive handedness}
\label{sec:cme_median_pos}

Once again, we start by injecting a spheromak CME with positive handedness (H=+1) in both solar wind backgrounds. For clarity, we will first focus on the 1D profiles, visible in Figure~\ref{fig:median_pos_1d_comp}. Similar to Figure~\ref{fig:extreme_pos_1d_comp}, the CME at maximum of activity arrives slightly in advance, but it could be due to the fact that the shock is more steep (especially visible in the velocity). What is surprising is that, contrary to the reference case where both CMEs had mostly positive $B_z$, in this case the CME at maximum of activity has a positive $B_z$ while the CME at minimum of activity has a clear negative component (end of day 3). Moreover, the ICME at minimum of activity has a longer ejecta, which means it takes more time to get back to a normal solar wind state after the crossing of the structure (the ejecta lasts 2.1 days, vs.\ 1.6 days at maximum of activity). This means that the median case is more geo-effective than the reference one. This is surprising, because the reference case was based on a CME event which has proven to be geo-effective by interaction with the wind background \citep{Hu2016, Marubashi2017, Gopalswamy2018, Scolini2019}, although it had been initially underestimated by the space-weather community \citep{Webb2017}.
This negative $B_z$ is also significant in magnitude, up to 15~nT. This is also surprising since the median CME is slower and less magnetized than the reference case. This highlights the fact that the geo-effectiveness of a CME does not depend only on the input CME parameters, but also on its interaction with its background and especially the HCS. 

\begin{figure}[th]
    \centering
    \plotone{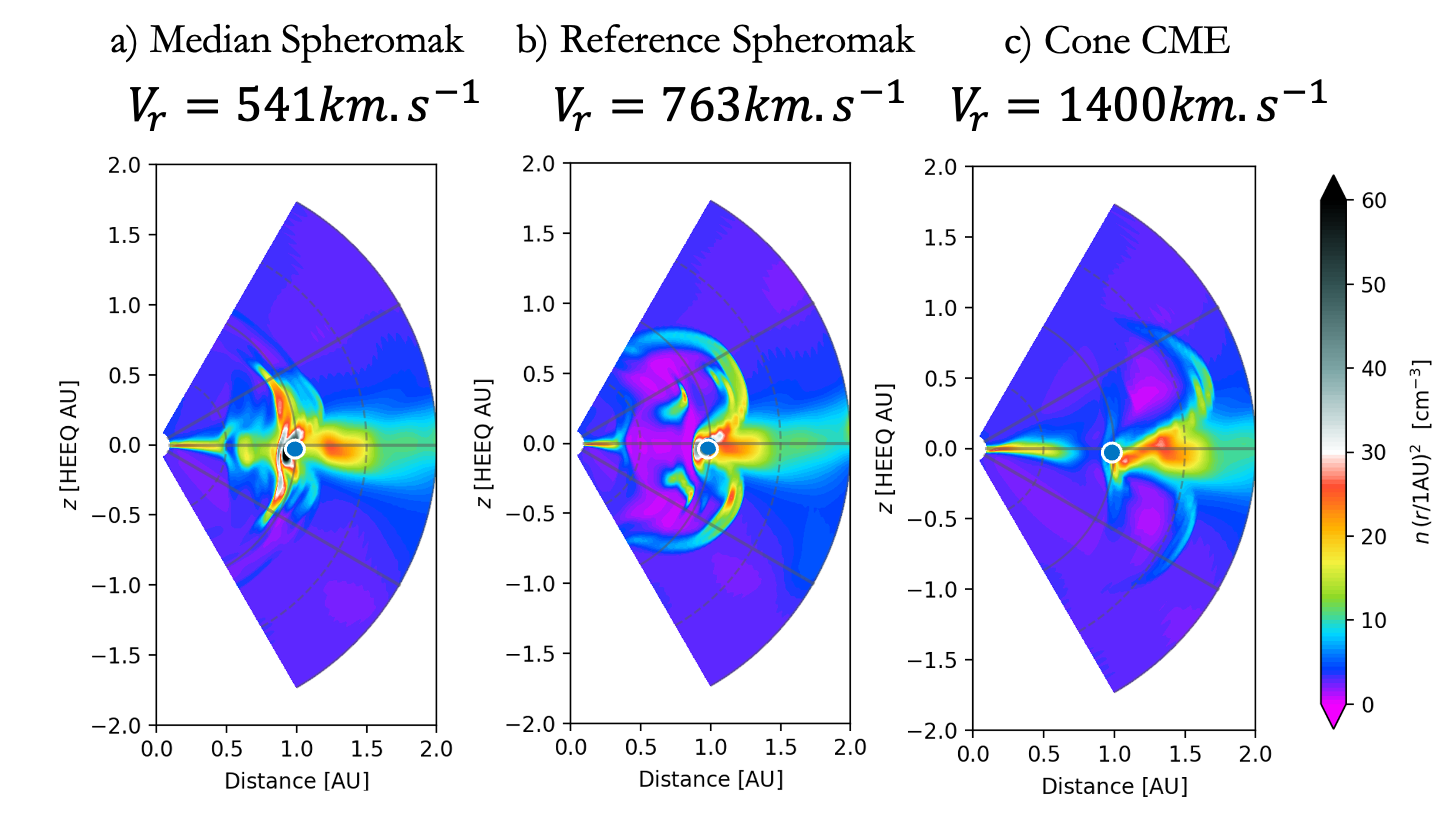}
    \caption{Comparison of the meridional view of the ICME at minimum of activity depending on the input radial speed. Each panel shows the meridional cut passing across Earth in the simulation (in the $x-z$ plane in HEEQ coordinates) when the ICME reaches Earth (blue point at 1~au). We show the density in order to visualize the magnetic ejecta as an under-density. The left panel shows the case for the median spheromak CME (Section \ref{sec:cme_median_pos}), the middle panel the case for the reference spheromak CME (Section \ref{sec:cme_12_july}) and the right panel the case for the cone CME (Section \ref{subsec:hydro_cme}). Each panel has the input CME speed at 0.1~au as a label. See Table~\ref{tab:cme_params} for a reminder of all the differences between the models. We see then that the faster the CME is, the more delay we observe between the equatorial and the polar parts of the ejecta, which means it is caused by the equatorial solar wind deceleration.}
    \label{fig:speed_slicing}
\end{figure}

\subsection{Slicing effect at minimum of activity}
\label{sec:cme_median_slicing}

Another interesting result we can derive from this study is a comparison of the ICME profiles at minimum of activity. We focus on this specific phase of the cycle because surprisingly, this is where we see the most differences between our cases. We would have expected the ICME at maximum of activity to show more disparity because of the complex wind background, but actually this complexity constrains the magnetic ejecta. As a result, the input parameters or even the modeling of the ICME (cone vs.\ spheromak) has little effect on the final profiles (in 2D or 1D, as can be seen in the previous figures), affecting only the size of the ejecta (by 0.1~au at most) but not its shape.

At minimum of activity however, the high structuring of the corona allows for more distinct effects that we can quantify. In Figure~\ref{fig:speed_slicing}, we compare the meridional profile for our three cases described in Table~\ref{tab:cme_params} at minimum of activity. From left to right, we show the median spheromak, the reference spheromak and the cone model. We show here the cuts in density to visualize the sheath as an over-density and the magnetic ejecta as an under-density.

We find again this slicing effect that we noticed before, where the high-latitude fast wind carries faster the northern and southern part of the ICME, while the equatorial part is slowed down by the slow equatorial wind. These three cases allow us to understand that this effect is not due to the modeling of the ICME, because we find it for both the cone and spheromak models (although the effect is reduced for the spheromak, probably because its internal magnetic structure makes it less sensitive to the wind background). The amount of slicing is highly sensitive to the input speed of the CME. The median and reference spheromaks have the same modeling, and yet the median case barely shows this effect. The major difference is the input speed: the spheromak case has an input speed closer to the fast wind, while the median case has an input speed closer to the slow wind. This indicates that this effect is not due so much to the acceleration caused by the fast wind, but rather by the slow-down caused by the slow equatorial wind. Then, as the ICME is faster, its equatorial part is much more slowed down, causing this geometric separation. The effect is even stronger for the cone model: the input speed appears to be even more important, but we recall that the definition of the input speed is slightly different for the cone model (as it does not have an expansion speed as in the spheromak model we also recall that the absence of an internal magnetic field makes the ICME more sensitive to the background wind.

This difference of behaviors in speed was also seen in ACE data in \cite{Regnault2020} for example. We know that at maximum of activity and during the declining phase, active regions have higher magnetic fluxes, which triggers more extreme events. But we could assume beyond this hypothesis that, even if an extreme event were to happen at minimum of activity, it would be slowed down drastically by the solar wind configuration and thus results in lower speed detected at Earth, just like what is seen in the ACE data (see also \cite{Chi2016} and \cite{Wu2016}). This result shows that the structuring of the wind itself has a strong influence on the ICME propagation, and not just the wind speed or specific structures at stream interfaces, and calls for more careful and realistic modeling of the solar corona.

\begin{figure}[th]
    \centering
    \plotone{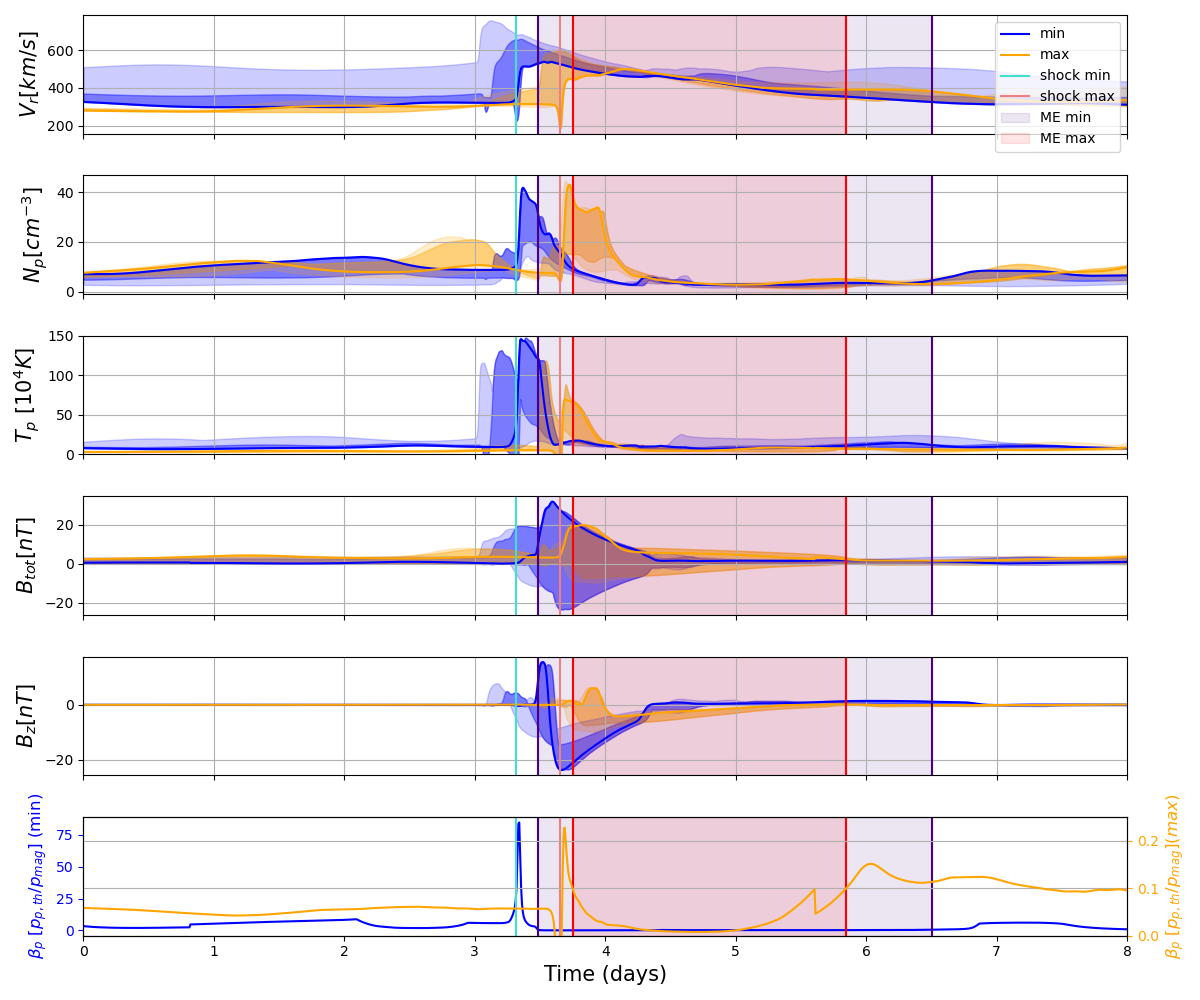}
    \caption{
    Comparison of the time evolution of the main physical quantities at 1~au at Earth position for the propagation of the median spheromak CME with negative handedness. The time is 
    the number of physical days that lasted the simulation. The blue line shows the minimum of activity case, while the orange line shows the maximum of activity case. From top to bottom, we show the radial velocity (in km/s), the proton number density (in $\rm{cm}^{-3}$), the temperature (in units of $10^4$ K), the total magnetic field (in nT), the $z$ components of the magnetic field in HEEQ coordinates (in nT) and the plasma beta parameter. The arrival of the shock is shown by a vertical line (light blue for minimum of activity case, red for maximum of activity case). The crossing of the ejecta is shown by a colored window (light purple for minimum of activity, light red for maximum of activity). We also show the vertical deviation at 5 and 10 degrees north and south of the ecliptic plane as colored shaded areas around the curves. Animated movies corresponding to this figure are available in the online version, showing the full propagation of the median spheromak CME with negative handedness in 2D cuts (equatorial, meridional and spherical) from 0.1~au to the Earth for both the minimum and maximum of activity backgrounds, with the evolution of the radial magnetic field, density and radial velocity.}
    \label{fig:median_neg_1d_comp}
\end{figure}

\begin{figure}[th]
    \centering
    \plotone{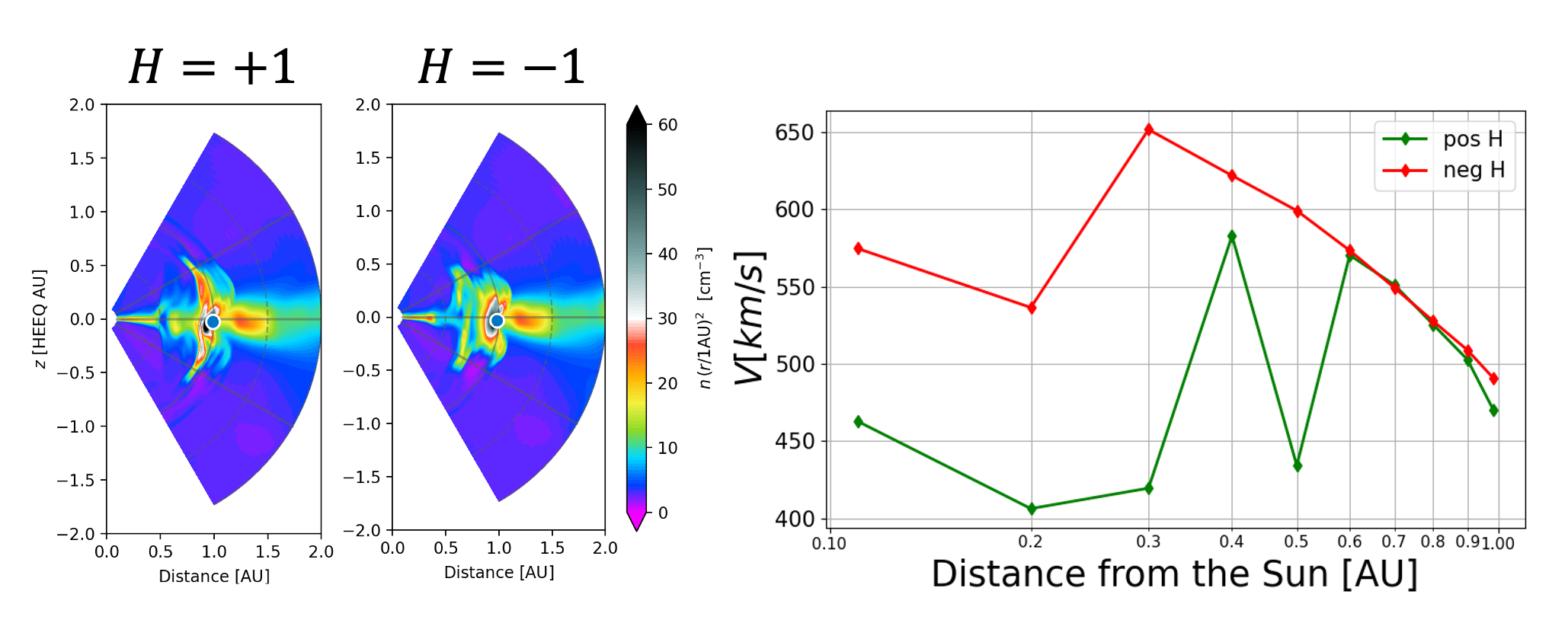}
    \caption{Comparison of the meridional profile of the ICME for positive (left panel) and negative (middle panel) handedness. We show the $x-z$ plane in HEEQ coordinates when the ICME reaches Earth (blue point at 1~au). We show the proton number density in $\rm{cm}^{-3}$. We can see that for the negative handedness case, the CME is elongated with a double sheath, probably due to the acceleration it gets from initial reconfiguration at 0.1~au. In the right panel, we show the radial evolution of the mean speed inside the ejecta for positive (green) and negative handedness (red). 
    }
    \label{fig:min_median_pos_neg}
\end{figure}

\subsection{Negative handedness}
\label{sec:cme_median_neg}

One final parameter study we undergo is to switch the handedness from positive to negative ($H=-1$) for the median case. In order to better compare all the cases and discuss our conclusions, we have summarized all the arrival times derived in the previous figures in table \ref{tab:timings}.

\begin{table}[]
    \centering
    \begin{tabular}{ | m{3cm} | m{2cm} | m{2cm} | m{2cm} | m{2cm} | m{2cm} | m{2cm} |}
        \hline
        CME Case & Activity level & Shock arrival time (days) & Ejecta arrival time (days) & Ejecta end time (days) & Sheath duration (hours) & Ejecta duration (days) \\ \hline \hline
        Cone & Minimum & $4.12$ & $4.28$ & $6.44$ & $3.83$ & $2.16$ \\ \hline
        Cone & Maximum & $3.78$ & $4.76$ & $7.25$ & $23.67$ & $2.49$ \\ \hline
        Reference & Minimum & $2.73$ & $3.07$ & $5.73$ & $8.17$ & $2.66$ \\ \hline
        Reference & Maximum & $2.67$ & $3.17$ & $6.36$ & $12.01$ & $3.19$ \\ \hline
        Median (positive H) & Minimum & $3.58$ & $3.82$ & $6.31$ & $5.67$ & $2.49$ \\ \hline
        Median (positive H) & Maximum & $3.51$ & $3.73$ & $5.39$ & $5.17$ & $1.66$ \\ \hline
        Median (negative H) & Minimum & $3.31$ & $3.49$ & $6.51$ & $4.17$ & $3.02$ \\ \hline
        Median (negative H) & Maximum & $3.65$ & $3.75$ & $5.85$ & $2.33$ & $2.10$ \\
        \hline
    \end{tabular}
    \caption{Table summarizing the key times found for each CME case and each activity level (minimum or maximum of activity). We indicate the shock arrival time (in days), the ejecta arrival time (in days), the ejecta end time (in days), and deduce from them the sheath duration (in hours) and the ejecta duration (in days).}
    \label{tab:timings}
\end{table}

We plot the 1D profiles at Earth in Figure~\ref{fig:median_neg_1d_comp}, with the same labels and visualization as before. This figure should be compared with Figure~\ref{fig:median_pos_1d_comp} to see the effect of the handedness. Here we can see a slight numerical artifact produced by EUHFORIA because of strong shocks where the velocity drops suddenly before rising again. It can disrupt locally certain physical quantities (like the $\beta_p$ plasma parameter), but this effect is purely local and does not impact the overall results and conclusions. At maximum of activity, the change of handedness causes the ICME to be slightly delayed at Earth, reaching 1~au after 3.65 days of propagation instead of 3.51  days for positive handedness. This can be explained by the fact that the ICME speed has been reduced (the speed at the shock is around 400 km/s vs.\ 500 km/s previously). The ejecta also lasts longer, with a return to solar wind parameters after 2.1 days instead of 1.66. More surprisingly, the ICME becomes slightly geo-effective, with a visible negative $B_z$ after day 4. However, these effects are minor compared to the minimum of activity case.

Once again, the case with the most disparity is surprisingly the case at minimum of activity. The most surprising feature is that the ICME at minimum of activity now arrives first, before the ICME at maximum of activity (this is the only case we have shown where this happens): the shock is move to day 3.31 instead of 3.58. The shock is also better defined, with a steepened slope. As for maximum of activity, the ICME also becomes more geo-effective: for negative handedness, the $B_z$ reaches -20~nT and stays negative, which would be conditions for a mild geomagnetic storm; whereas for positive handedness, the $B_z$ component would reach only -15~nT, but immediately go back to positive afterwards.

The above result is interesting, because it shows several properties of this case: the handedness can affect the ICME arrival time as well as its geo-effectiveness; these effects are more visible at minimum of activity than at maximum. One explanation we have for this phenomenon is shown in Figure~\ref{fig:min_median_pos_neg}. We show the meridional profile of the ICME hitting Earth for positive (left panel) and negative handedness (middle panel). For positive helicity, the ICME structure is typical, with an expanding spherical sheath followed by the magnetic ejecta. But for negative helicity, the sheath is more concentrated towards the equator, while the magnetic ejecta structure is not so visible. It almost seems that there is a second sheath forming behind the main one, with an overextended perturbation of the ejecta in latitude.

Although the two injected CMEs have the same input speed, the one with negative helicity is accelerated immediately after injection (at only 0.11~au, for an injection at 0.1~au), as seen in the right panel of Figure~\ref{fig:min_median_pos_neg}. This suggests that the rapid reconfiguration of the spheromak after injection to adjust to the wind background leads in the negative handedness case to the separation of the ICME, between an accelerated first part and then a decelerated one. Since we only changed the magnetic structure, we assume that this is the result of reconnection effects with the heliospheric magnetic field.
That would also explain why this effect is more visible at minimum of activity, which would be because the current sheet is closer to the equator, which means closer to the injection point as well as the propagation path of the ICME.  

Such effects of the impact of the handedness have been previously investigated by \cite{Chane2005, Chane2006}. They have indeed showed that the handedness itself can change the geo-effectiveness of the ICME as well as its arrival time, but only for a purely dipolar background configuration. Our results seem to extend this result even for more realistic backgrounds: at minimum of activity, the background is still dipolar enough so that the result can apply, but at maximum of activity the background becomes too complex and the influence of the handedness becomes less dominant. It is however difficult to generalize completely this result, as the CME in our case is only injected at 0.1~au, thus missing the propagation in the lower corona; for a more accurate result and better comparison, we would need to incorporate this phase as well, such as in \cite{Talpeanu2022}.

For verification, we have performed the same inversion of handedness for the reference spheromak case. We have found similar effects, although reduced compared to the median case (as expected due to the more intense internal magnetic field that makes the ICME less sensitive to the background). We retrieved the acceleration of the ICME at minimum of activity, although the internal separation was more difficult to see. This means that our results do apply to more extreme cases that can be seen at Earth. 
Finally, since the spheromak model is not connected to the Sun, it is more likely than other models to self-reconfigure, especially magnetically. It would be interesting to see if we could reproduce the same result with a CME model that has legs connected back to the Sun such as Fri3D \citep{Isavnin2016, Maharana2022}, Gibson \& Low model \citep{Gibson1998}, or the modified Titov-Démoulin model \citep{Titov2014, Regnault2023, Linan2023}.

\section{Discussion and Conclusion}
\label{sec:conclusion}

In this study, we investigate the role of the solar cycle on the propagation of ICMEs using numerical simulations. To do so, we start with a theoretical study that has an exploratory purpose. We select two dates that were representative of solar minimum (15th of December 2008) and solar maximum (20th of March 2015), based on previous studies. Then we use synoptic maps (GONG and GONG-ADAPT respectively) to drive the EUHFORIA model \citep{Pomoell2018} to compute the corresponding state of the heliosphere. Finally, we inject the same ICME within these two backgrounds, and quantify the differences and their origins, in order to better understand what to expect from a propagation at minimum versus a propagation at maximum of activity. We use several modelings for the ICME in order to test the robustness of our results: first a cone model to check for the hydrodynamical effects, then a linear force-free spheromak model to see the effect of an internal magnetic field. We also use parameters that were based on a true event (the CME that caused the geomagnetic storm on the 12ht of July 2012), as well as parameters derived to obtain a median ICME (based on ACE observations and EUHFORIA scaling, see \cite{Regnault2020} and \cite{Scolini2021_radial}).

We showed that the solar wind backgrounds selected were yielding similar speeds at Earth, but with very different structures. At minimum of activity, because the magnetic field of the Sun is mostly dipolar, the inner heliosphere is very organized, with slow wind at the equator and fast wind at the poles, while the HCS is 
close to the ecliptic plane. At maximum of activity on the contrary, coronal holes are more frequent at lower latitudes, leading to slow and fast wind everywhere, as well as a magnetic field with more complex structures and more polarity reversals. 

With a purely hydrodynamical ICME, we observe that the 1D speed profile at Earth is very similar, though the 3D structure of the ICME is very different. At minimum, the ICME is a direct hit at Earth, although its core is being slowed down by the slow equatorial wind, while the rest is accelerated by the fast polar wind. At maximum, the ICME is a flank hit, due to a northern-oriented deflection caused by a fast wind stream originating from a coronal hole. The ICME at maximum of activity arrives first at Earth with a 10-hour lead.
These geometrical results remain true for a magnetized ICME, although the inner magnetic field allows for the ICME to suffer less deformation and much less drag  along the propagation since the ICME reaches 1~au faster with a lower initial velocity. 
In this case, we can compare the geo-effectiveness of the two ICMEs by checking how negative their $B_z$ component is at Earth. For a positive handedness (H=+1), none are geo-effective (positive $B_z$), but for a negative handedness (H=-1), it is the ICME at minimum of activity which is actually the most geo-effective (-25 vs.\ -20~nT). The ICME at maximum of activity still arrives first at Earth, but only with a 2-hour lead.
These results remain true even for a median ICME. In this case, the difference in geo-effectiveness is even larger (-25 vs.\ -5~nT). This could explain why fast halo CMEs observed in solar maximum activity (2002) were poorly geoeffective \citep{Schmieder2020}. There is also a more distinct difference in arrival time: for a positive handedness, the ICME at maximum of activity arrives first with a 3-hour lead; with a negative handedness, it is the ICME at minimum of activity that arrives first with a 8-hour lead. 
We recover results obtained in \cite{Chane2006} for the minimum of activity, but show that the influence of the CME handedness is less dominant at maximum of activity. This could affect forecasts as it suggests that providing the handedness of the CME is only crucial at minimum of activity. This seems to be due to the self-reconfiguration of the ICME at injection, influenced by the dipolar magnetic background which accelerates the most favorable initial condition.
Another interesting result we obtained is that the deformation of the ICME at minimum of activity, caused by the structuring of the solar wind, depends on the speed of the ICME: the faster it is, the more important this effect will be, because the core of the ICME will be decelerated down even more noticeably.

In conclusion, we have shown that the same ICME will propagate very differently during solar minimum and maximum. The main factors are the organization of the solar wind, that can cause slow-downs or accelerations, but also the organization of the heliospheric magnetic field, that can cause  magnetic reconnection so allowing the overtaken plasma to be less stack in front of the ICME. 
In the cases studied, the ICME at minimum of activity was often the most geo-effective, which shows that the most powerful events will not necessarily happen at maximum of activity. This reinforces the need to quantify in the most precise way the coronal holes locations to anticipate deflections, as well as the HCS position and the ICME handedness to anticipate reconnection effects.

This study is a first step towards better understanding and quantifying the impact of the solar cycle on ICME propagation. This will become more and more important, as solar cycle 25 is on the rise, and space-weather forecasting facilities aim at delivering more and more reliable forecasts. To reach this goal, there are of course many ways to widen the scope of this study. A first natural step would be to include more intermediate states along a solar cycle (for example, following the previous cycle number 24) and see how the same ICME propagates. This would allow to better identify the key features that alter the propagation and when they change. Also, we could use other CME models (Fri3D, Gibson \& Low, etc.\citep{Maharana2022,Linan2023}), as well as other heliospheric models (ENLIL...) \citep{Odstrcil2004}to test even further the robustness of our results. 
We could also focus on reproducing specific events to validate our understanding of the key features of the heliosphere and their interaction with the ICME. Recent solar missions (such as PSP and Solar Orbiter) as well as future ones (such as PUNCH and Vigil) will even provide more data to add better constraints on the features of the inner heliosphere and the observed ICMEs. Finally, an important step would be to include the CME initialization inside the solar corona, in order to take into account the impact of the structures close to the Sun on the early propagation of the CME (similar to \cite{Lynch2022} for example).

\begin{acknowledgments}
The authors would like to thank Anwesha Maharana and Camilla Scolini for interesting discussions and important feedback.
This project has also received funding from the European Union’s Horizon 2020 research and innovation program under grant agreement No.~870405 (EUHFORIA 2.0) and the ESA project "Heliospheric modelling techniques“ (Contract No. 4000133080/20/NL/CRS). SP acknowledges support via the projects
C14/19/089  (C1 project Internal Funds KU Leuven), G.0B58.23N  (FWO-Vlaanderen), SIDC Data Exploitation (ESA Prodex-12), and Belspo project B2/191/P1/SWiM.
The resources and services used in this work were provided by the VSC (Flemish Supercomputer Centre), funded by the Research Foundation - Flanders (FWO) and the Flemish Government.
Data were acquired by GONG instruments operated by NISP/NSO/AURA/NSF with contribution from NOAA.
This work utilizes data produced collaboratively between AFRL/ADAPT and NSO/NISP.
We recognise the collaborative and open nature of knowledge creation and dissemination, under the control of the academic community as expressed by Camille No\^{u}s at \url{http://www.cogitamus.fr/indexen.html}.
\end{acknowledgments}

\appendix

\section{Determination of the CME shock, sheath and magnetic ejecta}
\label{appendix:shock_sheath_mc}

An ICME exhibits different substructures that are essential to distinguish, due to the different underlying physics in each one of them. In this study, we need to be able to distinguish between the CME-driven shock, the CME sheath and the magnetic ejecta (also called magnetic cloud in some studies, e.g.\ when the presence of a flux-rope can be confirmed). To do so, we need to determine the shock time $t_{shock}$ (which is the beginning of the sheath), the end of the sheath and thus beginning of the magnetic ejecta time $t_{in}$, and the end of the magnetic ejecta time $t_{out}$. 

\begin{figure}[th!]
    \centering
    \plotone{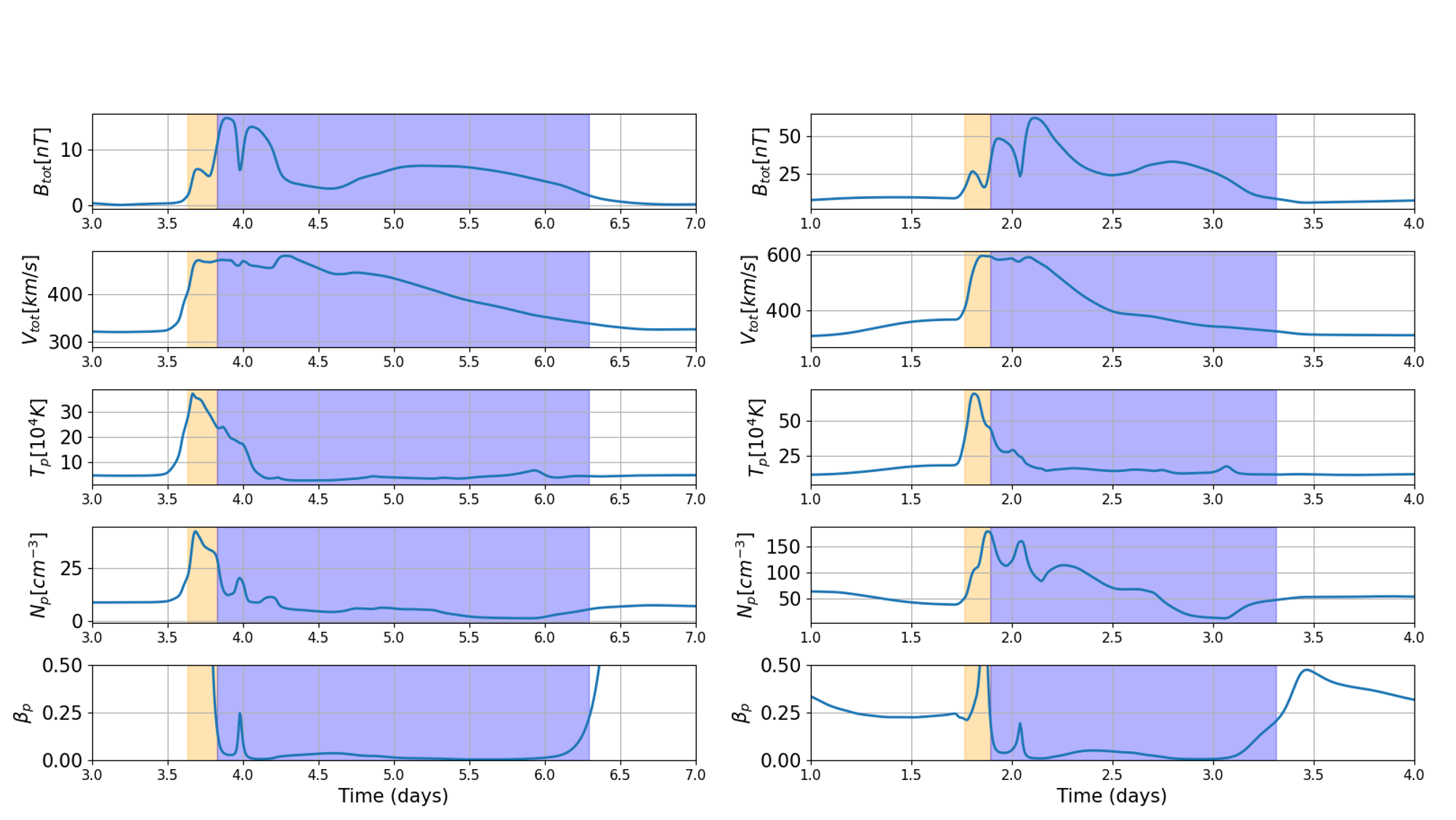}
    \caption{
    Examples of detection of the shock, sheath and magnetic cloud for various cases. For each case, the first row is the total magnetic amplitude in nT, the second row the total velocity amplitude in km/s, the third row the temperature in $10^4$ K, the fourth row the number density for the protons in $\rm{cm}^{-3}$ and the last row $\beta$ parameter of the plasma. The left panel shows the median case at minimum of activity at 1~au. The right panel shows the same case, but at 0.4~au instead of 1~au. A yellow rectangle highlights the region corresponding to the sheath, and a blue rectangle the region corresponding to the magnetic cloud.}
    \label{fig:shock_sheath_mc_criteria}
\end{figure}

To determine $t_{shock}$ at various distances for various case, we use a modified version of the criterion used in \cite{Scolini2021_radial}:
\begin{equation}
    \left(v(t_i) - v(t_i - \Delta t) \geqslant v_{thresh}\right) \quad OR \quad \left(\frac{n(t_i)}{n(t_i-\Delta t)} \geqslant n_{thresh}\right) \quad OR \quad \left(\frac{B(t_i)}{B(t_i-\Delta t)} \geqslant B_{thresh}\right),
\end{equation}
where $t_i$ is a generic time in the time series, $\Delta t$ is a time delay to compare to the steady wind state before the event
, and $v_{thresh}$, $n_{thresh}$ and $B_{thresh}$ are threshold parameters to identify the shock. \cite{Scolini2021_radial} set their own thresholds to the following values: $v_{thresh}=20$ km/s, $n_{thresh}=1.2$, $B_{thresh}=1.2$. However, these were for a comparison between the CME run and the corresponding wind-only simulation, which means they used a wind model for reference. In our case, we use only one CME run and compare present time with previous time data. Their values were also optimized for a specific wind background set at maximum of activity. This means that our threshold values need to be different.
We have thus adjusted these values by trying different combinations, and selected the most robust and efficient ones. These final parameter values are: $\Delta t=10$ (with 10 minutes between each output, this means an interval of around 1.6 hours), $v_{thresh}=30$ km/s, $n_{thresh}=1.5$, $B_{thresh}=1.5$
($n_{thresh}$ and $B_{thresh}$ do not have units because they are ratios). Since we have a minimum of activity configuration, the HCS is close to the equatorial place and as a result, the magnetic field can become very close to 0 locally, producing false detection of the shock because of these low values. To avoid this, we have set an additional threshold of 1 G for the local magnetic field. On top of the automatic selection given by this criterion, systematic visual verification has been made to ensure the validity of the results.

For $t_{in}$, we use a criterion from the $\beta$ plasma. This criterion is adapted from in situ solar wind measurements \citep{Lepping2005} and has already been used in \cite{Scolini2021_radial}. In the observations, the sheath ends (and the magnetic ejecta begins) when $\beta_{p,obs} \leqslant 0.3$. For EUHFORIA simulations, the threshold can range from 0.1 to 1. \cite{Scolini2021_radial} found that the value of 0.5 yielded good results in their cases. 
For our cases, it is the value of 0.1 which usually gives the best results. However, we sometimes had to adjust it manually in order to get a ME selection consistent with the other physical quantities. For clarity, we have indicated the value of $\beta$ used for the border selection in table \ref{tab:beta_values}.

\begin{table}[]
    \centering
    \begin{tabular}{|c|c|c|c|}
        \hline
        CME Case & Activity level & $\beta$ threshold & Corresponding Figure \\ \hline \hline
        Cone & Minimum & $4.0$ & \ref{fig:cone_1d_comp} \\ \hline
        Cone & Maximum & $0.15$ & \ref{fig:cone_1d_comp} \\ \hline
        Reference & Minimum & $0.1$ & \ref{fig:extreme_pos_1d_comp} \\ \hline
        Reference & Maximum & $0.1$ & \ref{fig:extreme_pos_1d_comp} \\ \hline
        Median (positive H) & Minimum & $0.25$ & \ref{fig:median_pos_1d_comp} \\ \hline
        Median (positive H) & Maximum & $0.1$ & \ref{fig:median_pos_1d_comp} \\ \hline
        Median (negative H) & Minimum & $0.5$ & \ref{fig:median_neg_1d_comp} \\ \hline
        Median (negative H) & Maximum & $0.1$ & \ref{fig:median_neg_1d_comp} \\ \hline
    \end{tabular}
    \caption{Table of the values of the $\beta$ parameter used as threshold for detection of the borders of the CME sheath.}
    \label{tab:beta_values}
\end{table}

$t_{out}$ is set when the $\beta$ parameter goes above the same threshold as for $t_{in}$. We include an exception for reconnection effects that can occur within the magnetic ejecta, and that usually generate perturbations of the $\beta$ for only a few points. The goal is to not take into account spikes in the $\beta$ parameter.

Figure~\ref{fig:shock_sheath_mc_criteria} shows how these various criteria manage to adapt to different cases to detect automatically all the substructures. The left panel shows the same case as the left panel in Figure~\ref{fig:cme_median_profiles}, which is the median CME at minimum of activity at 1~au. 
There is some reconnection happening within the magnetic ejecta, because the $\beta$ spikes above 0.2 at December 18 around noon. However the magnetic ejecta is not over yet, and our criterion can adapt to these cases. The right panel shows the same case, but at 0.4~au instead of 1~au. 
Adjustment of $\Delta t$ can be needed for artificial satellites closer to the injection point, because the shock will steepen with the distance, we usually then divide it by 2.

\begin{figure}[th]
    \centering
    \plotone{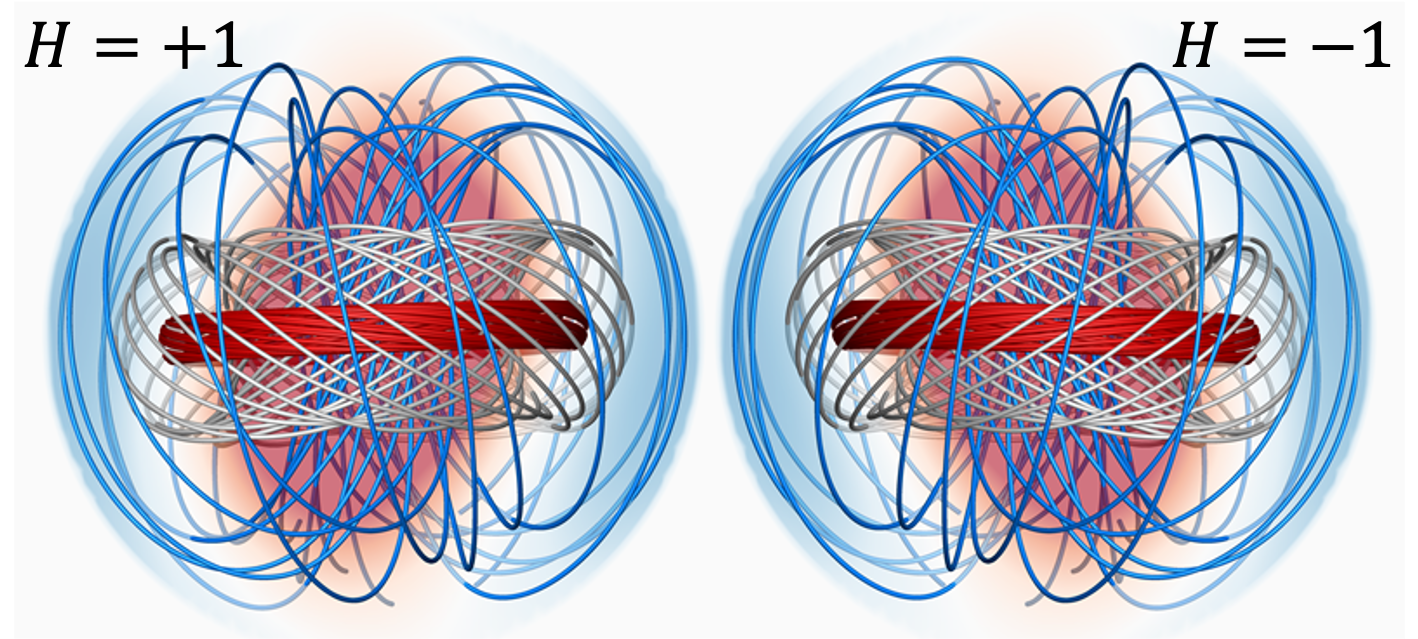}
    \caption{
    3D representation of the magnetic configuration of the initial spheromak. We represent three sets of magnetic field lines: the blue ones correspond to the outer edge of the spheromak, the red ones to the axis of the internal flux rope, and the white ones to the orientation of the flux rope field lines. In the left panel, the handedness is positive since the flux-rope corresponds to a right-handed helix. In the right panel, the handedness is negative since the flux-rope corresponds to a left-handed helix. Credits: Camilla Scolini.
    }
    \label{fig:handedness}
\end{figure}

\section{Handedness of a spheromak}
\label{appendix:handedness}

The magnetic field in the spheromak model is expressed as follows (in the local spherical coordinate $(r,\theta,\phi)$ frame in which the origin is the center of the spheromak): 
\begin{align}
    B_r & = 2B_0\frac{j_1\left(\alpha r\right)}{\alpha r}\rm{cos}\,\theta, \\
    B_\theta & = -B_0\left[\frac{j_1\left(\alpha r\right)}{\alpha r} + \partial_r j_1\left(\alpha r\right)\right]\rm{sin}\,\theta, \\
    B_\phi & = H B_0 j_1\left(\alpha r\right)\rm{sin}\,\theta,
\end{align}
where $B_0$ is a parameter determining the magnetic field strength, $j_1(x)$ is the spherical Bessel function of order one, $\alpha$ is chosen so that 
$\alpha r_0$ is the first zero of $j_1(x)$, which yields $\alpha r_0 \approx 4.4934$ 
($r_0$ is the radius of the spheromak), and $H$ is the handedness. 

The handedness is a dimensionless parameter that can be equal to either +1 or -1. 
Only the azimuthal component of the magnetic field is affected by the handedness. In Figure~\ref{fig:handedness}, we show a 3D representation of the resulting spheromak magnetic field (in the meridional plane) for positive and negative handedness. It can be linked to the sign of the magnetic helicity of the originating active region, which is conserved over time \citep{Berger2005}. The magnetic helicity quantifies how much the magnetic field is sheared and twisted. The handedness only retains its sign. 
Spheromaks with a positive handedness ($H=+1$) exhibit a right-handed helix flux-rope, while spheromaks with a negative handedness ($H=-1$) exhibit a left-handed helix flux-rope (shown by the torus located in the central part of the spheromak). Handedness can be estimated using empirical relationships such as hemispheric rules \citep[majority of positive handedness in the southern hemisphere and negative handedness in the northern hemisphere, ][]{Bothmer1998, Pevtsov2008}, or by analyzing morphological features or the photospheric magnetic field of the active region or/and erupting filament \citep{Demoulin2009, Palmerio2017}.

There is sometimes a confusion with other names for quantities that have the same or a similar function. The name "chirality" is used interchangeably with handedness (as for example in \cite{Shiota2016} or \cite{Scolini2019}). Confusion may arise when the handedness is referred to as a "helicity parameter", as in \cite{Jin2017}. However the handedness is not the helicity, it is only its sign. 

\bibliography{cme_solar}{}
\bibliographystyle{aasjournal}

\end{document}